\documentclass[nonacm,sigconf,natbib=true]{acmart}
\usepackage[norelsize,ruled,vlined,linesnumbered]{algorithm2e}
\usepackage{algorithmic}
\SetKwComment{Comment}{$\triangleright$\ }{}
\usepackage{subfigure}
\usepackage{booktabs}
\AtBeginDocument{%
  }

\newtheoremstyle{mystyle}% % Name
    {1.5mm}%	% Space above
    {1.5mm}% 	% Space below
    {\it}%		% Body font
    {0.0mm}%	% Indent amount
    {\scshape}% % Theorem head font
    {.}%	% Punctuation after theorem head
    { }%	% Space after theorem head, ' ', or \newline
    {}%		% Theorem head spec (can be left empty, meaning `normal')
\theoremstyle{mystyle}
\newtheorem{definition}{Definition}
\newtheorem{example}{Example}

\newtheorem{theorem}{Theorem}

\newtheorem{remark}{Remark}

\newcommand{\wsq}{\hspace{\fill}$\square$}

\newcommand{\vs}{\vspace{1.5mm}}

\setcopyright{acmlicensed}
\copyrightyear{2018}
\acmYear{2018}
\acmDOI{XXXXXXX.XXXXXXX}

\setcopyright{acmcopyright}
\copyrightyear{}
\acmYear{}
\acmDOI{}
\acmConference[]{}{}{}
\acmBooktitle{}
\acmPrice{}
\acmISBN{}

\setcopyright{none}
\renewcommand\footnotetextcopyrightpermission[1]{} % removes footnote with conference information in first column
\acmConference[Conference acronym 'XX]{Make sure to enter the correct conference title from your rights confirmation emai}{June 03--05, 2018}{Woodstock, NY}
\acmISBN{978-1-4503-XXXX-X/18/06}

\begin{document}

%%
%% The "title" command has an optional parameter,
%% allowing the author to define a "short title" to be used in page headers.
\title{How to Mine Potentially Popular Items? A Reverse MIPS-based Approach}

\author{Daichi Amagata}
\authornote{Both authors contributed equally to this research. Daichi Amagata is the corresponding author.}
\email{amagata.daichi@ist.osaka-u.ac.jp}
% \orcid{1234-5678-9012}
\author{Kazuyoshi Aoyama}
\authornotemark[1]
\email{aoyama.kazuyoshi@ist.osaka-u.ac.jp}
\affiliation{%
  \institution{The University of Osaka}
  \city{Suita, Osaka}
  % \state{Ohio}
  \country{Japan}
}
\author{Keito Kido}
\email{kido.keito@ist.osaka-u.ac.jp}
\affiliation{%
  \institution{The University of Osaka}
  \city{Suita, Osaka}
  \country{Japan}
}
\author{Sumio Fujita}
\email{sufujita@lycorp.co.jp}
\affiliation{%
  \institution{LY Corporation}
  \city{Chiyoda, Tokyo}
  \country{Japan}
}

%%
%% By default, the full list of authors will be used in the page
%% headers. Often, this list is too long, and will overlap
%% other information printed in the page headers. This command allows
%% the author to define a more concise list
%% of authors' names for this purpose.
\renewcommand{\shortauthors}{Aoyama et al.}

%%
%% The abstract is a short summary of the work to be presented in the
%% article.
\begin{abstract}
The $k$-MIPS ($k$ Maximum Inner Product Search) problem has been employed in many fields.
Recently, its reverse version, the reverse $k$-MIPS problem, has been proposed.
Given an item vector (i.e., query), it retrieves all user vectors such that their $k$-MIPS results contain the item vector.
Consider the cardinality of a reverse $k$-MIPS result.
A large cardinality means that the item is potentially popular, because it is included in the $k$-MIPS results of many users.
This mining is important in recommender systems, market analysis, and new item development.
Motivated by this, we formulate a new problem.
In this problem, the score of each item is defined as the cardinality of its reverse $k$-MIPS result, and the $N$ items with the highest score are retrieved.
A straightforward approach is to compute the scores of all items, but this is clearly prohibitive for large numbers of users and items.
We remove this inefficiency issue and propose a fast algorithm for this problem.
Because the main bottleneck of the problem is to compute the score of each item, we devise a new upper-bounding technique that is specific to our problem and filters unnecessary score computations.
We conduct extensive experiments on real datasets and show the superiority of our algorithm over competitors. 
\end{abstract}

%%
%% The code below is generated by the tool at http://dl.acm.org/ccs.cfm.
%% Please copy and paste the code instead of the example below.
%%
% \begin{CCSXML}
% <ccs2012>
%    <concept>
%        <concept_id>10002951.10003317.10003338.10003346</concept_id>
%        <concept_desc>Information systems~Top-k retrieval in databases</concept_desc>
%        <concept_significance>500</concept_significance>
%        </concept>
%  </ccs2012>
% \end{CCSXML}

% \ccsdesc[500]{Information systems~Top-k retrieval in databases}

% \keywords{inner product search, high-dimensional vector, algorithm}

% \received{20 February 2007}
% \received[revised]{12 March 2009}
% \received[accepted]{5 June 2009}

\maketitle

\section{Introduction}  \label{sec:introduction}
Due to the proliferation of machine learning techniques, many applications embed objects, such as users and items, into $d$-dimensional vectors.
To measure the relevance of these vectors, inner products are usually used.
Furthermore, $k$ Maximum Inner Product Search ($k$-MIPS) and its variants (e.g., \cite{aoyama2023simpler,hirata2022solving,hirata2022cardinality,hirata2023categorical,nakama2021approximate}) are used in many fields, including information retrieval \cite{xiang2019accelerating,zhang2021joint}, data mining \cite{pham2021simple,tan2021norm}, database \cite{abuzaid2019index,li2017fexipro}, artificial intelligence \cite{liu2020understanding,liu2019bandit}, machine learning \cite{guo2020accelerating,yu2017greedy}, and recommender systems \cite{wang2020dnn,bachrach2014speeding}.
For example, in recommender systems, a recommendation list for a given user is often obtained via $k$-MIPS, where a query vector is the user vector \cite{li2023semantic,li2024mutual}.

\vs
\noindent
\textbf{Motivation.}
Notice that $k$-MIPS is usually user-centric, as queries are user vectors.
To provide an item-centric analysis, the reverse $k$-MIPS problem was proposed in \cite{amagata2021reverse,amagata2023reverse}.
Given an item vector (as a query) and $k$, this problem retrieves all user vectors whose $k$-MIPS results contain the query item vector.
This problem is useful, for example, for effective item advertisements, as it identifies users who would prefer the items.
This concept is employed in recommender systems: for example, \cite{li2023will} shows the effectiveness of the reverse search.

Now it is important to notice that the cardinality of a reverse $k$-MIPS result implies the ``potential popularity'' of a given item.
This is because a large cardinality implies that many users would prefer a given item.
Therefore, analyzing items with large reverse $k$-MIPS result sizes suggests which items will be popular next.
This is important in market analysis, developing new items, marketing optimization, and (global) purchase pattern prediction.
This motivates us to address a new problem of retrieving the top-$N$ items with the largest reverse $k$-MIPS result size.
(A formal definition appears in Section \ref{sec:preliminary}.)

The output of this problem is different from the so-called most popular items (i.e., items with the largest number of interactions).
Table \ref{tab:top-popular} compares the top-5 results of the most popular items and our problem, and no common items exist in the results.
This is because our problem considers ``potentially'' popular (or high-ranked) items, as can be seen from the context of $k$-MIPS (i.e., items in $k$-MIPS results have large \textit{predicted} inner products with given query vectors).
On the other hand, the most popular items are not necessarily ranked in the top-$k$ for given users.

\vs
\noindent
\textbf{Challenge.}
Consider that we are given a set $\mathbf{U}$ of $n$ user vectors and a set $\mathbf{P}$ of $m$ item vectors.
We assume that these vectors are $d$-dimensional ones.
Let the score of an item vector be the cardinality of the reverse $k$-MIPS result of the item.
A straightforward algorithm that solves our problem is to run a state-of-the-art $k$-MIPS algorithm for each user in $\mathbf{U}$ to obtain the $k$-MIPS results of all users.
From these $k$-MIPS results, given an item, we can compute how many users contain the item in their $k$-MIPS results.
Although this algorithm can obtain the correct answer, it is time consuming, since (i) it needs to run $n$ $k$-MIPS operations and (ii) it computes the scores of all items included in the $k$-MIPS results.
It is easy to imagine that applications want to test multiple values of $N$ and $k$, thus this time-consuming approach is infeasible.
Another intuitive algorithm is to run a state-of-the-art reverse $k$-MIPS algorithm for each item in $\mathbf{P}$ to compute its score.
This algorithm also suffers from the same computational inefficiency as that of the straightforward algorithm.

\begin{table}[t]
    \centering
    \caption{Most popular vs. reverse 10-MIPS result size-based order on MovieLens.
    Their results are described by movie IDs.}
    \label{tab:top-popular}
    \begin{tabular}{ccc}   \toprule
        Order   & Most popular  & Reverse 10-MIPS result size   \\ \midrule
        1       & 318           & 201821                        \\
        2       & 296           & 140369                        \\
        3       & 356           & 106115                        \\
        4       & 593           & 177325                        \\
        5       & 2571          & 87798                         \\ \bottomrule
    \end{tabular}
\end{table}

Now we have a question:
Do we always need to run (reverse) $k$-MIPS for each user (item) vector to obtain the correct answer?
In other words, can we filter unnecessary (reverse) $k$-MIPS without losing correctness?
To provide the output of our problem interactively, we need to design an efficient algorithm that gives a positive answer to this question.
However, this is a non-trivial task.

\vs
\noindent
\textbf{Contribution.}
In this work, we provide a positive answer to the above question by proposing an efficient and exact algorithm for our problem.
In pre-processing, we compute an upper-bound score of each item in an efficient manner (by assuming that there is a maximum value of $k$).
This offline algorithm, which is done only once, takes at most $O(nm)$ time.
It is important to notice that this time complexity does not have a factor of $d$.
This result \textit{theoretically} implies that our offline algorithm obtains an upper-bound score of each item (as if) without inner product computations.
The upper-bound scores (i) guide us to the item vectors whose exact scores should be computed earlier and (ii) filter unnecessary score computations.
Our online algorithm is designed based on this rationale.

To summarize, our contributions in this work are as follows:
\begin{itemize}
    \setlength{\leftskip}{-5.0mm}
    \item   We formulate a new problem that outputs $N$ items with the largest reverse $k$-MIPS result size.
    \item   We propose an offline algorithm that efficiently computes a tight upper-bound score of each item vector.
            Then, we propose an exact online algorithm that can filter items that do not contribute to the final result.
            We show that our algorithm theoretically cannot be beaten by straightforward algorithms.
    \item   We conduct extensive experiments on real-world datasets.
            The results demonstrate that \textit{our algorithm is at least two orders of magnitude faster than competitors}.
\end{itemize}

The rest of this paper is organized as follows.
Section \ref{sec:preliminary} introduces preliminary information including problem definition and baselines.
We review related works in Section \ref{sec:related-work}.
Our algorithm is presented in Section \ref{sec:proposal}, and experimental results are reported in Section \ref{sec:experiment}.
This paper is concluded in Section \ref{sec:conclusion}.

\section{Preliminary}   \label{sec:preliminary}

\subsection{Problem Definition}
Let $\mathbf{U}$ be a set of $n$ user vectors.
Also, let $\mathbf{P}$ be a set of $m$ item vectors.
We assume that these vectors are high-dimensional ones, and the dimensionality is denoted by $d$.
Therefore, for example, a user vector $\mathbf{u}$ is represented as $(u_1, u_2, ..., u_d)$.
In addition, as long as the relevance between a user vector $\mathbf{u} \in \mathbf{U}$ and an item vector $\mathbf{p} \in \mathbf{P}$ is measured by their inner product $\mathbf{u} \cdot \mathbf{p}$, our algorithm is orthogonal to the ways of generating $\mathbf{U}$ and $\mathbf{P}$.
Note that inner product is a standard measurement in embedding-based retrieval systems \cite{chang2020pre,aouali2022reward,lin2022proposed,doan2023asymmetric}.
For ease of presentation, we first define the $k$ maximum inner product search ($k$-MIPS) problem.

\begin{definition}[\textsc{$k$-MIPS problem}]
Given a user vector $\mathbf{u}$ as a query, a set $\mathbf{P}$ of item vectors, and $k$, this problem returns a set $\mathbf{S}_{k}(\mathbf{u})$ of item vectors in $\mathbf{P}$ such that $|\mathbf{S}_{k}(\mathbf{u})| = k$ and for each $\mathbf{p} \in\mathbf{S}_{k}(\mathbf{u})$ and $\mathbf{p}' \in \mathbf{P} \backslash \mathbf{S}_{k}(\mathbf{u})$, $\mathbf{u} \cdot \mathbf{p} \geq \mathbf{u} \cdot \mathbf{p}'$.
(Ties are broken arbitrarily.)
\end{definition}

\noindent
The above problem finds the item vectors that have the largest inner product with a given user (query) vector. 
We next introduce the reverse $k$-MIPS problem \cite{amagata2021reverse,amagata2023reverse}.

\begin{definition}[\textsc{Reverse $k$-MIPS problem}]
Given an item vector $\mathbf{p}$ as a query, a set $\mathbf{U}$ of user vectors, a set $\mathbf{P}$ of item vectors, and $k$, this problem returns a set $\mathbf{T}_{k}(\mathbf{p})$ such that 
\begin{equation}
    \mathbf{T}_{k}(\mathbf{p}) = \{\mathbf{u} \,|\, \mathbf{u} \in \mathbf{U}, \mathbf{p} \in \mathbf{S}_{k}(\mathbf{u})\}. \label{eq:rmips}
\end{equation}
\end{definition}

\vs
\noindent
The reverse $k$-MIPS problem finds the user vectors whose $k$-MIPS results contain a given query item vector $\mathbf{p}$.
Now we are ready to introduce our problem\footnote{A similar setting is considered in \cite{vlachou2010identifying}, but it assumes low-dimensional Euclidean space, which is totally different from our vector setting.}.

\begin{definition}[\textsc{Top-$N$ item search based on reverse $k$-MIPS result size}]  \label{definition:problem}
Given a set $\mathbf{U}$ of user vectors, a set $\mathbf{P}$ of item vectors, $k$, and a result size $N$, this problem returns a set $\mathbf{R}$ such that $|\mathbf{R}| = N$ and for each $\mathbf{p} \in \mathbf{R}$ and $\mathbf{p'} \in \mathbf{P} \backslash \mathbf{R}$, $|\mathbf{T}_{k}(\mathbf{p})| \geq |\mathbf{T}_{k}(\mathbf{p'})|$.
(Ties are broken arbitrarily.)
\end{definition}

Our problem uses $|\mathbf{T}_{k}(\mathbf{p})|$ as the \textit{score} of $\mathbf{p}$.
Hereinafter, let $score_{k}(\mathbf{p}) = |\mathbf{T}_{k}(\mathbf{p})|$.
Table \ref{tab:notation} summarizes the notations frequently used in this paper.
The objective of this work is to develop a fast and \textit{exact} algorithm for the problem in Definition \ref{definition:problem}.
As with the existing reverse $k$-MIPS studies \cite{amagata2021reverse,amagata2023reverse,huang2023sah}, we assume that there is a maximum value of $k$, denoted by $k_{max}$, and $k_{max} = O(1)$.

\begin{table}[t]
    \centering
    \caption{Summary of notations}
    \label{tab:notation}
    \begin{tabular}{cl}   \toprule
        Notation                                                & Meaning                                           \\ \midrule
        $\mathbf{u}$ ($\mathbf{U}$)                             & $d$-dim. user vector (set of $n$ user vectors)    \\
        $\mathbf{p}$ ($\mathbf{P}$)                             & $d$-dim. item vector (set of $m$ item vectors)    \\
        $\mathbf{u} \cdot \mathbf{p}$                           & inner product of $\mathbf{u}$ and $\mathbf{p}$    \\
        $k$                                                     & result size of MIPS                               \\
        $\mathbf{S}_{k}(\mathbf{u})$                            & result set of $k$-MIPS of $\mathbf{u}$            \\
        $\mathbf{T}_{k}(\mathbf{p})$                            & result set of reverse $k$-MIPS of $\mathbf{p}$    \\
        $N$                                                     & result size of our problem                        \\
        $|\mathbf{T}_{k}(\mathbf{p})| = score_{k}(\mathbf{p})$  & score of $\mathbf{p}$                             \\
        $uscore_{k}(\mathbf{p})$                                & upper-bound of $score_{k}(\mathbf{p})$            \\
        $A_{i}$                                                 & array storing (approx.) $k$-MIPS result of $\mathbf{u}_{i}$   \\
        \bottomrule
    \end{tabular}
\end{table}

\subsection{Baseline}   \label{sec:preliminary:baseline}
A straightforward algorithm for our problem is to use a state-of-the-art exact $k$-MIPS algorithm.
Specifically, for each $\mathbf{u}_{i} \in \mathbf{U}$,
\begin{enumerate}
    \setlength{\leftskip}{-3.0mm}
    \item   we run the $k$-MIPS algorithm to obtain $\mathbf{S}_{k}(\mathbf{u})$.
    \item   For each $\mathbf{p} \in \mathbf{S}_{k}(\mathbf{u})$, we increment $score_{k}(\mathbf{p})$ by one.
            (Assume that $score_{k}(\mathbf{p}) = 0$ at initialization.)
\end{enumerate}
From Equation (\ref{eq:rmips}), it is trivial that the above algorithm computes the exact score of each item in $\mathbf{P}$.
Then, at last, we return $N$ items with the largest score.

Another baseline algorithm is to use a state-of-the-art exact reverse $k$-MIPS algorithm.
This is more intuitive: for each $\mathbf{p}_{j} \in \mathbf{P}$, we run the reverse $k$-MIPS algorithm to obtain $|\mathbf{T}_{k}(\mathbf{p}_{j})|$.
Then, $N$ items with the largest score are returned.

The main computational bottleneck of our problem is to compute the score of each item.
The two baseline algorithms suffer from this issue severely, as they compute the exact score of each item.
Our algorithm alleviates this issue by filtering unnecessary score computations.

\section{Related Work}  \label{sec:related-work}

\subsection{Exact $k$-MIPS Algorithm}
Existing exact $k$-MIPS algorithms can be categorized into two approaches: space-partitioning and linear scan.
The state-of-the-art space-partitioning algorithms \cite{curtin2013fast} employ Cone-tree \cite{ram2012maximum}.
Unfortunately, this approach suffers from the curse of dimensionality, and its performance deteriorates significantly for large $d$.

To avoid performance degradation owing to the curse of dimensionality, efficient algorithms based on linear scan were proposed in \cite{li2017fexipro,teflioudi2015lemp,teflioudi2016exact}.
They show that linear scan-based algorithms are much faster than space-partitioning algorithms.
Motivated by this observation, we also employ a linear scan-based approach when running $k$-MIPS-related operations.

\subsection{Approximate $k$-MIPS Algorithm}
To speed up $k$-MIPS while sacrificing result quality, approximate $k$-MIPS algorithms were devised.
There are four approaches in these works: sampling, locality-sensitive hashing (LSH), quantization, and proximity graph.
Sampling-based algorithms \cite{liu2019bandit,ding2019fast} cannot deal with vectors containing negative values, so they lose generality.
LSH-based algorithms \cite{yan2018norm,song2021promips} can enjoy theoretical accuracy and performance guarantees but are slow in practice.
Due to the success of ScaNN \cite{guo2020accelerating}, quantization-based algorithms have been receiving attention \cite{zhang2021joint,zhang2023query}.
The proximity graph-based approach also has a great attention \cite{tan2021norm,feng2023reinforcement}.
Although these approaches yield good empirical performances, their search result quality is not controllable and the worst time is still $O(m)$.

In this paper, we do not consider these algorithms, because they cannot control the output quality of our problem.
Specifically, if we use one of the approximation algorithms for $k$-MIPS, we obtain approximate $\mathbf{S}(\cdot)$.
This also leads to an incorrect reverse $k$-MIPS result, which can be seen from Equation (\ref{eq:rmips}), and it loses the correct score of $\mathbf{p}$.
Then, in this approximation case, items with large exact scores tend to disappear from the top-$N$ result, whereas items with small exact scores can appear in the result.
This is not desirable for the applications introduced in Section \ref{sec:introduction}.
We hence focus on an exact solution.

\subsection{Existing Reverse $k$-MIPS Algorithm}
The reverse $k$-MIPS problem was proposed in \cite{amagata2021reverse,amagata2023reverse}\footnote{As pointed out in \cite{amagata2021reverse,amagata2023reverse,huang2023sah}, although a similar concept was considered in \cite{vlachou2010reverse,vlachou2011monochromatic,vlachou2013branch}, the techniques in these papers are not available for the reverse $k$-MIPS problem.
This is because they assume a different data space and extremely low $d$ (e.g., $d < 10$).}.
Simpfer \cite{amagata2021reverse} is an exact reverse $k$-MIPS algorithm.
Its rationale is to address the $k$-MIPS decision problem, that is, it answers whether a given query item is included in the $k$-MIPS result of a given user.
Simpfer employs some data structures that can solve the $k$-MIPS decision problem in a constant time in the best case.
We borrow this idea when we need to compute the exact score of a given item.

In \cite{amagata2023reverse}, an approximation version of reverse $k$-MIPS is considered to address a question: can approximation improve the search efficiency?
Although the answer is positive, it shows that the improvement is only slight.
In \cite{huang2023sah}, SAH, a different approximation algorithm, was proposed.
SAH employs an LSH-based approach and achieves a sub-linear time to $m$, but it requires a sub-quadratic space to $n$, which does not scale well to a large number of users---a main requirement of information retrieval systems.
Again, we do not consider these approximation algorithms, as they do not (i) provide $\mathbf{T(\cdot)}$ correctly and (ii) have any error bound for $\mathbf{T(\cdot)}$.

A problem similar to reverse $k$-MIPS is addressed in \cite{amagata2025approximate,bian2024qsrp}.
However, because its output is different from that of the reverse $k$-MIPS problem, their techniques are not available.

\section{Proposed Algorithm}    \label{sec:proposal}
This section presents our algorithm that solves the problem in Definition \ref{definition:problem} exactly and efficiently.
In an offline phase, we build some data structures (e.g., arrays) that are used to filter unnecessary score computations online.
This pre-processing is done only once.
Given $N$ and $k$, our algorithm computes the $N$ item vectors with the largest score.

\subsection{Main Idea and Rationale}    \label{sec:proposal:main-idea}
The main bottleneck of our problem is to compute the score of a given item vector.
The baseline algorithms compute the score of each item vector, so they suffer from a huge computational cost.
To avoid this inefficiency, we employ an upper-bounding technique.
Let $uscore_{k}(\mathbf{p})$ be an upper-bound of $score_{k}(\mathbf{p})$, i.e., $uscore_{k}(\mathbf{p}) \geq score_{k}(\mathbf{p})$.
Furthermore, let $\tau$ be a threshold, which corresponds to the (intermediate) $N$-th largest score.
If $\tau \geq uscore_{k}(\mathbf{p})$, $score_{k}(\mathbf{p})$ cannot be larger than $\tau$, thus we can safely filter computing $score_{k}(\mathbf{p})$.

In pre-processing, we compute $uscore_{k}(\mathbf{p})$ for each $\mathbf{p} \in \mathbf{P}$ and for each $k \in [1,k_{max}]$.
To maximize this filtering efficiency, we should have $uscore_{k}(\mathbf{p}) \approx score_{k}(\mathbf{p})$.
In addition, $uscore_{k}(\mathbf{p})$ should be computed efficiently (e.g., linear time to $n$ and $m$).
To satisfy these requirements, we devise a new technique for efficiently computing tight $uscore_{k}(\mathbf{\cdot})$.
We first use only $O(k_{max}) = O(1)$ item vectors.
For each user vector $\mathbf{u}_{i} \in \mathbf{U}$, by accessing these $O(k_{max})$ item vectors, we obtain an approximate $k_{max}$-MIPS result.
We store this result set in an array $A_{i}$, and $A_{i}^{k}$ is used to denote the $k$-th largest inner product for $\mathbf{u}_{i}$ among the $O(k_{max})$ item vectors.
From $A_{i}^{k}$, we know which item vectors are possibly included in $\mathbf{S}_{k}(\mathbf{u}_{i})$ (the exact $k$-MIPS result of $\mathbf{u}_{i}$) and cannot be in $\mathbf{S}_{k}(\mathbf{u}_{i})$.
For an item vector $\mathbf{p}$, the ``possibly included in'' case means that $\mathbf{T}_{k}(\mathbf{p})$ may contain $\mathbf{u}_{i}$.
In this case, we increment $uscore_{k}(\mathbf{p})$ by one, and $uscore_{k}(\mathbf{p})$ is obtained in this way.

More specifically, we use Cauchy–Schwarz inequality
\begin{equation}
    \mathbf{u}_{i} \cdot \mathbf{p} \leq \|\mathbf{u}_{i}\|\|\mathbf{p}\|   \label{eq:cauchy}
\end{equation}
to evaluate whether $\mathbf{p}$ can be in $\mathbf{S}_{k}(\mathbf{u}_{i})$ or not, where $\|\mathbf{p}\|$ is the $l_{2}$ norm of $\mathbf{p}$.
If $\|\mathbf{u}_{i}\|\|\mathbf{p}\| < A^{k}_{i}$, $\mathbf{p}$ cannot be in $\mathbf{S}_{k}(\mathbf{u}_{i})$.
Therefore, for each $\mathbf{p} \in \mathbf{P}$ such that $\|\mathbf{u}_{i}\|\|\mathbf{p}\| \geq A^{k}_{i}$, $uscore_{k}(\mathbf{p})$ is incremented by one.

In an online (i.e., a query) phase, given $N$ and $k$, we first sort the item vectors in $\mathbf{P}$ in descending order of $uscore_{k}(\cdot)$.
We run a reverse $k$-MIPS in this order.
Assume that we now access the $i$-th item vector $\mathbf{p}$.
If $\tau \geq uscore_{k}(\mathbf{p})$, $\mathbf{p}$ and the subsequent item vectors cannot be in the top-$N$ result.
Hence, we can safely filter score computation for these item vectors, significantly speeding up the query processing efficiency.

\subsection{Pre-processing}
We elaborate on how to construct our data structure and compute upper-bound scores.
Recall that this pre-processing is done only once, and our data structure is available for arbitrary $k \in [1, k_{max}]$.
From now on, we assume that the item vectors in $\mathbf{P}$ are sorted in descending order of norm.

An overview of our pre-processing was described in Section \ref{sec:proposal:main-idea}.
We make this pre-processing efficient with two optimization approaches.
The first employs \textit{increment pruning} \cite{teflioudi2015lemp}.
The second one is derived from an observation: \textit{the number of item vectors accessed to obtain $\mathbf{S}_{k_{max}}(\mathbf{\cdot})$ is different for each user vector.}
To understand this observation, consider a linear scan of $\mathbf{P}$ to obtain $\mathbf{S}_{k_{max}}(\mathbf{\mathbf{u}_{i}})$.
We can early stop the scan when $A^{k_{max}}_{i} \geq \|\mathbf{u}_{i}\|\|\mathbf{p}\|$, which is seen from Inequality (\ref{eq:cauchy}).
As each user vector differs from the others, its norm can also be different from those of the others, leading to the observation.
(Example \ref{example:mips} will further detail this observation.)
Now assume that we have a budget \textit{for inner product computations} (to make this pre-processing efficient).
Recall that we want to make $uscore_{k}(\mathbf{p})$ nearly $score_{k}(\mathbf{p})$.
This can be achieved by obtaining $\mathbf{S}_{k}(\cdot)$ or $\mathbf{S}_{k_{max}}(\cdot)$ for all user vectors, but this is impractical under the limited budget.
From the above observation, it can be seen that a uniform budget assignment for each user vector $\mathbf{u}_{i}$ is not effective in obtaining an (approximate but) accurate $\mathbf{S}_{k_{max}}(\mathbf{u}_{i})$.
Therefore, we have to solve the non-trivial challenge of assigning an appropriate number of item vectors to a given user vector.
This is an important task for obtaining an accurate $\mathbf{S}_{k_{max}}(\cdot)$ for many users (i.e., to obtain tight upper-bound scores) under the budget constraint.
Our second approach is to address this challenge, which is specific to our problem, by dynamic assignment.

\subsubsection{\textbf{Incremental Pruning}}
To employ this technique, we divide each user/item vector into two vectors.
Let $d' \in [1,d]$ be the dimension for the vector partition, and $\mathbf{u} = \mathbf{u}^{l}\mathbf{u}^{r}$, where $\mathbf{u}^{l} = (u_1, ..., u_{d'})$ and $\mathbf{u}^{r} = (u_{d'+1}, ..., u_{d})$.
Then, we have
\begin{equation}
    \mathbf{u}_{i} \cdot \mathbf{p} \leq \mathbf{u}^{l}_{i} \cdot \mathbf{p}^{l} + \|\mathbf{u}^{r}_{i}\|\|\mathbf{p}^{r}\|.   \label{eq:inc}
\end{equation}
By setting $d' = O(1)$, this evaluation is also done in $O(1)$ time (after computing each $\|\mathbf{u}^{r}\|$ and $\|\mathbf{p}^{r}\|$).
Notice that this bound is tighter than the one obtained from Inequality (\ref{eq:cauchy}).
In our pre-processing, this bound is also used to make $uscore_{k}(\cdot)$ tight while keeping efficiency.

\subsubsection{\textbf{Dynamic Number Assignment}}
Let $B$ be a budget, which is the number of inner product computations for all users in pre-processing.
We set $B = O(nk_{max})$.
A straightforward assignment of $B$ for each user is a uniform assignment, i.e., each user can use $O(k_{max})$ items.
However, as noted earlier, this is ineffective, so we consider a dynamic assignment.

Let $B = B_{1} + B_{2}$.
First, we assign $B_{1} / n$ to each user vector.
Each user vector $\mathbf{u}_{i}$ can scan the first $B_{1} / n$ item vectors and stores the intermediate $k_{max}$-MIPS result in $A_{i}$.
Inequality (\ref{eq:cauchy}) suggests that, for some user vectors, say $\mathbf{u}$, the intermediate $k_{max}$-MIPS result is not guaranteed to be $\mathbf{S}_{k_{max}}(\mathbf{u})$.

\begin{figure}[t]
   \centering
   \includegraphics[width=0.75\linewidth]{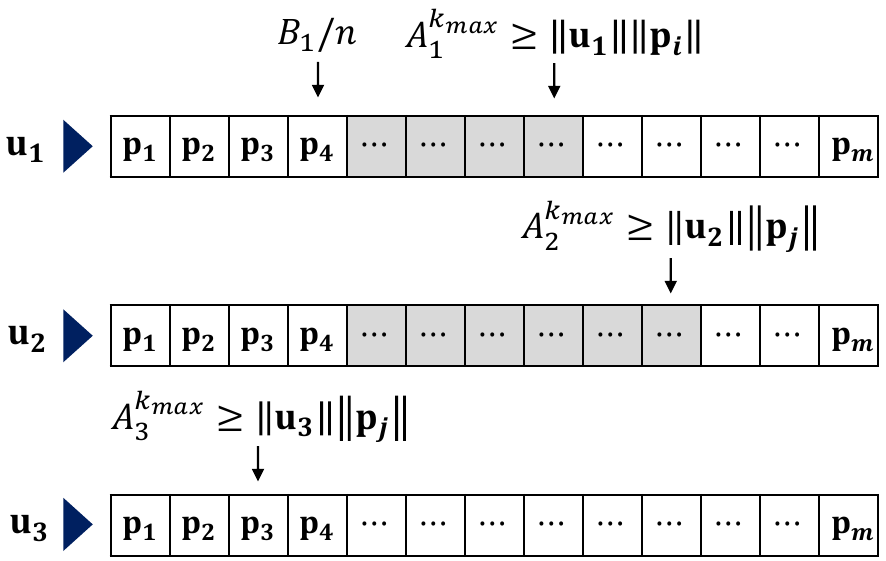}
   \caption{Illustration of Example \ref{example:mips}}
   \label{fig:example}
\end{figure}

\begin{example} \label{example:mips}
We assume that we have $\|\mathbf{p}_{i}\| \geq \|\mathbf{p}_{i+1}\|$ for each $i \in [1, n-1]$.
In addition, $B_{1} / n = 4$ and $k_{max} = 2$.
Figure \ref{fig:example} illustrates $\mathbf{u}_{1}$, $\mathbf{u}_{2}$, $\mathbf{u}_{3}$, and $\mathbf{P}$.
In the case of $\mathbf{u}_{1}$, by scanning the first four item vectors, we obtain $A_{1}^{k_{max}}$, but $A_{1}^{k_{max}}$ is not guaranteed to be $\mathbf{S}_{k_{max}}(\mathbf{u}_{1})$.
To guarantee it, $\mathbf{u}_{1}$ needs to scan to the $i$-th item vector.
(The gray parts show the desirable additional budget for each user vector.)
Similarly, $\mathbf{u}_{2}$ needs to scan to the $j$-th item vector to guarantee $\mathbf{S}_{k_{max}}(\mathbf{u}_{2})$.
On the other hand, $\mathbf{u}_{3}$ needs to scan only the first two item vectors to guarantee $\mathbf{S}_{k_{max}}(\mathbf{u}_{3})$.
\end{example}

\begin{figure}[!t]
    \begin{center}
        \subfigure[MovieLens]{%
		\includegraphics[width=0.485\linewidth]{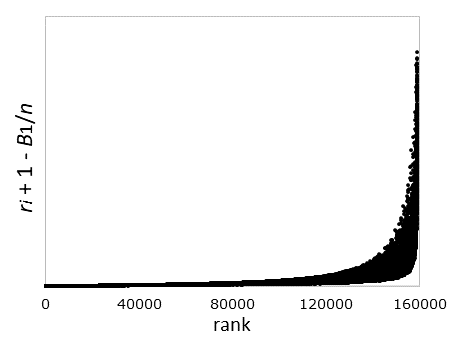} \label{fig:function:movielens}}
        \subfigure[Netflix]{%
		\includegraphics[width=0.485\linewidth]{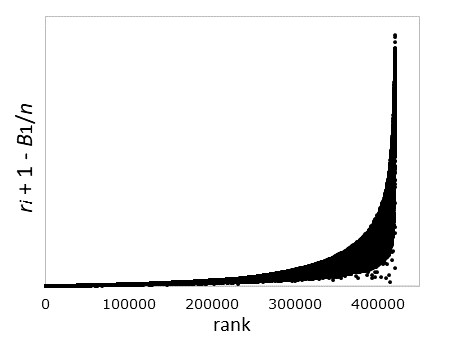}   \label{fig:function:netflix}}
        \caption{Relationship between $x_{i}$ and ($r_{i} + 1 - B_{1} / n$)}
        \label{fig:function}
    \end{center}
\end{figure}

\noindent
We see, from this example, that each user vector $\mathbf{u}_{i} \in \mathbf{U}$ needs to scan to the $r_{i}$-th item vector to guarantee $\mathbf{S}_{k_{max}}(\mathbf{u}_{i})$ (if $r_{i} > B_{1} / n$), where $r_{i}$ is variable.
Therefore, $\mathbf{u}_{i}$ needs additional $(r_{i} + 1 - B_{1} / n)$ item vector accesses.
We need to assign the remaining budget based on this number (variable).

To enable effective assignment, we consider a function that outputs a budget for a given $\mathbf{u}_{i}$.
To obtain an \textit{easy} input for this function, we sort the user vectors, such that $r_{i} > B_{1} / n$, in ascending order of $(r_{i} + 1 - B_{1} / n)$.
Based on this sort order, each user vector has a rank, e.g., the first rank has the smallest $(r_{i} + 1 - B_{1} / n)$.
Let $x_{i}$ be the rank of $\mathbf{u}_{i}$.
We map $\langle x_{i}, r_{i} + 1 - B_{1} / n\rangle$ into a 2-dimensional space.
Figure \ref{fig:function} depicts the result of this mapping on MovieLens and Netflix datasets.
(The other datasets we used in our experiments show similar shapes.)
From this observation, we use the following exponential function:
\begin{equation}
    f(x) = \alpha \exp(\beta x) + \gamma,    \label{eq:f_x}
\end{equation}
where $\alpha$, $\beta$, and, $\gamma$ are parameters.
We set $\alpha = O(1)$ and $\gamma= O(1)$.
Let $\mathbf{U}'$ be a set of user vectors whose $k_{max}$-MIPS results are not guaranteed to be $\mathbf{S}_{k_{max}}(\cdot)$ under the  $B_{1} / n$ budget.
We determine $\beta$ by solving
\begin{equation}
    \int_0^{|\mathbf{U}'|}f(x)dx = \frac{\alpha(\exp(\beta |\mathbf{U}'|) - 1)}{\beta} + \gamma |\mathbf{U}'| \approx B_{2}.   \label{eq:dynamic}
\end{equation}
Then, each $\mathbf{u}_{i} \in \mathbf{U}'$ can have a budget of scanning $\mathbf{P}$ to the $f(x_{i})$-th item vector.

\begin{figure}[!t]
    \begin{center}
        \subfigure[MovieLens]{%
		\includegraphics[width=0.485\linewidth]{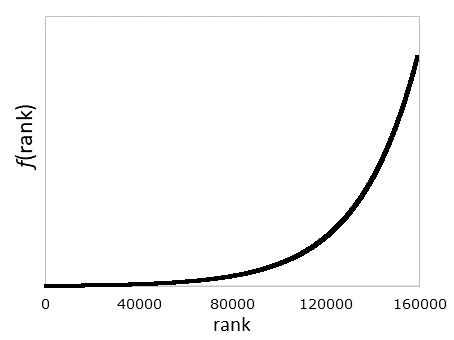} \label{fig:function:movielens_approx}}
        \subfigure[Netflix]{%
		\includegraphics[width=0.485\linewidth]{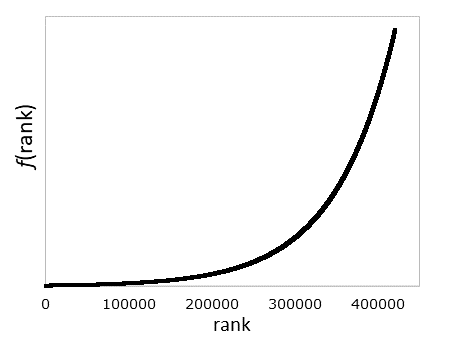}   \label{fig:function:netflix_approx}}
        \caption{Concrete examples of the result of Equation (\ref{eq:f_x})}
        \label{fig:function_approx}
    \end{center}
\end{figure}

The function in Equation (\ref{eq:f_x}) has three merits.
First, it approximates the real distribution well.
Figure \ref{fig:function_approx} illustrates concrete examples of Equation (\ref{eq:f_x}) on MovieLens and Netflix (after solving Equation (\ref{eq:dynamic})).
Comparing Figure \ref{fig:function} with Figure \ref{fig:function_approx}, we see that Equation (\ref{eq:f_x}) works well.
Second, no training is required.
The last merit is $O(1)$ time for computing $f(x)$.

\subsubsection{\textbf{Pre-processing Algorithm}}
Based on the above techniques, we compute $uscore_{k}(\mathbf{p})$ for each $\mathbf{p} \in \mathbf{P}$ and $k \in [1, k_{max}]$.
Our pre-processing algorithm has the following five steps.
Algorithm \ref{algo:pre-processing} details this pre-processing.

\begin{algorithm}[!t]
    \caption{\textsc{Pre-processing}}
    \label{algo:pre-processing}
    \DontPrintSemicolon
    \KwIn {$\mathbf{U}$, $\mathbf{P}$, and $k_{max}$}

    \textbf{foreach} $\mathbf{u}_{i} \in \mathbf{U}$ and $\mathbf{p}_{j} \in \mathbf{P}$ \textbf{do} Compute $\|\mathbf{u}_{i}\|$ and $\|\mathbf{p}_{j}\|$\;            \label{algo:pre-processing:norm_b}
    Sort $\mathbf{P}$ in descending order of norm\;                                                                                                                     \label{algo:pre-processing:norm_e}
    SVD transformation of $\mathbf{U}$ and $\mathbf{P}$, $\mathbf{U'} \gets \varnothing$, $\mathbf{U''} \gets \varnothing$\;                                            \label{algo:pre-processing:svd_b}
    \textbf{foreach} $\mathbf{u}_{i} \in \mathbf{U}$ and $\mathbf{p}_{j} \in \mathbf{P}$ \textbf{do} Compute $\|\mathbf{u}^{r}_{i}\|$ and $\|\mathbf{p}^{r}_{j}\|$\;    \label{algo:pre-processing:svd_e}
    \ForEach {$\mathbf{u}_{i} \in \mathbf{U}$}{                                                                                                                         \label{algo:pre-processing:uniform_b}
        $\theta \leftarrow 0$   \Comment*[r]{\scriptsize threshold initialization}
        \ForEach {$j \in [1,B_{1} / n]$}{
            \eIf {$|\|\mathbf{u}_{i}\|\|\mathbf{p}_{j}\| > \theta$}{                                                                                                    \label{algo:pre-processing:mips_b}
                \If {$\mathbf{u}^{l}_{i} \cdot \mathbf{p}^{l}_{j} + |\|\mathbf{u}^{r}_{i}\|\|\mathbf{p}^{r}_{j}\| > \theta$}{                                           \label{algo:pre-processing:update_b}
                    Compute $\mathbf{u}_{i} \cdot \mathbf{p}_{j}$\;
                    Update $A_{i}$ and $\theta$ if necessary                                                                                                            \label{algo:pre-processing:update_e}
                }
            }{
                \textbf{break}                                                                                                                                          \label{algo:pre-processing:mips_e}
            }
        }
        \If {$\theta < \|\mathbf{u}_{i}\|\|\mathbf{p}_{j}\|$}{
            Compute $j'$ such that $\theta \geq \|\mathbf{u}_{i}\|\|\mathbf{p}_{j'}\|$\;                                                                                \label{algo:pre-processing:bs}
            Add $\langle \mathbf{u}_{i}, j' + 1 - B_{1} / n\rangle$ into $\mathbf{U'}$\;
        }
    }
    Sort $\mathbf{U'}$ in ascending order of $j + 1 - \frac{B_{1}}{n}$\;
    Determine $\beta$ in Equation (\ref{eq:f_x}) by Equation (\ref{eq:dynamic}), $B' \gets 0$\;                                                                         \label{algo:pre-processing:uniform_e}
    \ForEach {$\mathbf{u}_{i} \in \mathbf{U'}$}{                                                                                                                        \label{algo:pre-processing:dynamic_b}
        $B_{2,i} \gets$ the budget determined by Equation (\ref{eq:f_x})\;
        \ForEach {$j \in [\frac{B_{1}}{n} + 1, \frac{B_{1}}{n} + 1 + B_{2,i} + B']$}{
            \eIf {$\|\mathbf{u}_{i}\|\|\mathbf{p}_{j}\| > A^{k_{max}}_{i}$}{
                Run lines \ref{algo:pre-processing:update_b}--\ref{algo:pre-processing:update_e}
            }{
                $B' \gets \frac{B_{1}}{n} + 1 + B_{2,i} + B' - j$\;
                \textbf{break}
            }
        }
        \textbf{if} $A^{k_{max}}_{i} < \|\mathbf{u}_{i}\|\|\mathbf{p}_{j}\|$ \textbf{then} $pos_{i} \gets j$, Add $\mathbf{u}_{i}$ into $\mathbf{U''}$                  \label{alg:pre-processing:record}
    }
    $uscore_{k}(\mathbf{p}_{i}) \gets 0$ for each $\mathbf{p}_{j} \in \mathbf{P}$ and $k \in [1, k_{max}]$\;                                                            \label{alg:pre-processing:ub_b}
    \ForEach {$\mathbf{u}_{i} \in \mathbf{U''}$}{
        $\lambda^{*}_{i} \gets 0$\;
        \ForEach {$j \in [pos_{i} + 1, m]$}{
            \textbf{if} $\theta \geq \|\mathbf{u}_{i}\|\|\mathbf{p}_{j}\|$ \textbf{then} \textbf{break}\;
            $\lambda \gets \mathbf{u}^{l}_{i} \cdot \mathbf{p}^{l}_{j} + \|\mathbf{u}^{r}_{i}\|\|\mathbf{p}^{r}_{j}\|$\;
            $\lambda^{*}_{i} \gets \max \{\lambda^{*}_{i}, \lambda\}$   \Comment*[r]{\scriptsize Equation (\ref{eq:ip_upper-bound})}
            \ForEach {$k \in [1,k_{max}]$}{
                \textbf{if} $\lambda \geq A_{i}^{k}$ \textbf{then} $uscore_{k}(\mathbf{p}_{i}) \gets uscore_{k}(\mathbf{p}_{i}) + 1$                                    \label{algo:pre-processing:ub}
            }
        }
    }
    \ForEach {$i \in [1, n]$ and $k \in [1, k_{max}]$}{
        \If {$\mathbf{p}_{j}$ is included in the first $k$ elements of $A_{i}$}{
            $uscore_{k}(\mathbf{p}_{j}) \gets uscore_{k}(\mathbf{p}_{j}) + 1$                                                                                           \label{alg:pre-processing:ub_e}
        }
    }
\end{algorithm}

\vs
\noindent
\underline{\textbf{(1) Norm computation} (lines \ref{algo:pre-processing:norm_b}--\ref{algo:pre-processing:norm_e}).}
For each user vector $\in \mathbf{U}$, we compute its norm.
We do the same for each item vector, then sort the item vectors in descending order of norm.
Without loss of generality, we assume that $\|\mathbf{p}_{i}\| \geq \|\mathbf{p}_{i+1}\|$ for each $i \in [1, n-1]$, i.e., $\mathbf{p}_{i}$ is the $i$-th vector in this sort order.

\vs
\noindent
\underline{\textbf{(2) SVD transformation} (lines \ref{algo:pre-processing:svd_b}--\ref{algo:pre-processing:svd_e}).}
Notice that $\mathbf{U}$ and $\mathbf{P}$ respectively are represented as a matrix, where each row and column correspond to a vector and dimension.
Therefore, we can apply SVD (Singular Value Decomposition) to $\mathbf{U}$ and $\mathbf{P}$, then compute $\|\mathbf{u}^{r}_{i}\|$ and $\|\mathbf{p}^{r}_{j}\|$ for each $\mathbf{u}_{i} \in \mathbf{U}$ and $\mathbf{p}_{j} \in \mathbf{P}$.

SVD corresponds to a global re-ordering of the $d$ dimensions \cite{feng2023reinforcement}, and after SVD, the first dimensions of each vector have higher absolute values than those of the remaining dimensions.
Therefore, we have $\mathbf{u}_{i} \cdot \mathbf{p} \approx \mathbf{u}_{i}^{l} \cdot \mathbf{p}^{l}$, suggesting that SVD can make the bound in Inequality (\ref{eq:inc}) tighter.

\vs
\noindent
\underline{\textbf{(3) Uniform budget assignment} (lines \ref{algo:pre-processing:uniform_b}--\ref{algo:pre-processing:uniform_e}).}
We assign $B_{1} / n$ to each user $\mathbf{u}_{i} \in \mathbf{U}$ as a budget.
For each $\mathbf{u}_{i} \in \mathbf{U}$, to obtain the initial $A_{i}$, we run a sequential scan of $\mathbf{P}$ to at most the $\frac{B_{1}}{n}$-th item vectors while applying Inequality (\ref{eq:inc}).
Note that whenever we have $A_{i}^{k_{max}} \geq \|\mathbf{u}_{i}\|\|\mathbf{p}\|$, we stop the scan, since the remaining item vectors cannot be in $\mathbf{S}_{k_{max}}(\mathbf{u}_{i})$.
In this early stop case, we have $A_{i} = \mathbf{S}_{k_{max}}(\mathbf{u}_{i})$.

On the other hand, if $\mathbf{u}_{i}$ does not have this case, $A_{i}$ is not guaranteed to be $\mathbf{S}_{k_{max}}(\mathbf{u}_{i})$.
We then search for the minimum $j$ such that $A_{i}^{k_{max}} \geq \|\mathbf{u}_{i}\|\|\mathbf{p}_{j}\|$, which can be efficiently done by a binary search.
We maintain such user vectors in $\mathbf{U}'$ and sort them in ascending order of $j + 1 - B_{1} / n$.
After that, we determine $\beta$ in Equation (\ref{eq:f_x}) via Equation (\ref{eq:dynamic}).

\vs
\noindent
\underline{\textbf{(4) Dynamic budget assignment} (lines \ref{algo:pre-processing:dynamic_b}--\ref{alg:pre-processing:record}).}
For each $\mathbf{u}_{i} \in \mathbf{U}'$, we assign its next budget via Equation (\ref{eq:f_x}).
Let $f(x_{i}) = B_{2,i}$.
We scan $\mathbf{P}$ from the $(\frac{B_{1}}{n} + 1)$-th item vector to at most the $(\frac{B_{1}}{n} + 1 + B_{2,i})$-th one.
During the scan, we update $A_{i}$.
If we have $A_{i} = \mathbf{S}_{k_{max}}(\mathbf{u}_{i})$ before consuming the assigned budget, we pool the unconsumed budget so that the next user vector can use the pooled budget.
On the other hand, if $\mathbf{u}_{i}$ does not guarantee that $A_{i} = \mathbf{S}_{k_{max}}(\mathbf{u}_{i})$, we record the last accessed position $pos_{i} = (\frac{B_{1}}{n} + 1 + B_{2,i})$ and add $\mathbf{u}_{i}$ into $\mathbf{U}''$.

\vs
\noindent
\underline{\textbf{(5) Upper-bound score computation} (lines \ref{alg:pre-processing:ub_b}--\ref{alg:pre-processing:ub_e}).}
For each user vector $\mathbf{u}_{i} \in \mathbf{U}''$, we have $pos_{i}$, and  $\mathbf{u}_{i}$ has not accessed item vector $\mathbf{p}_{j}$ such that $j \in [pos_{i}+1,m]$.
For such $\mathbf{p}_{j}$, if we have
\begin{equation}
    \mathbf{u}^{l}_{i} \cdot \mathbf{p}^{l}_{j} + \|\mathbf{u}^{r}_{i}\|\|\mathbf{p}^{r}_{j}\| \geq A_{i}^{k}   \label{eq:upper-bound}
\end{equation}
for a given $k \in [1, k_{max}]$, $\mathbf{p}_{j}$ may be included in $\mathbf{S}_{k}(\mathbf{u_{i}})$.
Therefore, we increment $uscore_{k}(\mathbf{p}_{j})$ by one.
For $\mathbf{u}_{i}$ this is done by scanning a set of $\mathbf{p}_{j}$ such that $j \in [pos_{i}+1,m]$ until we have $A^{k_{max}}_{i} \geq \|\mathbf{u}_{i}\|\|\mathbf{p}_{j}\|$.
In addition, we record 
\begin{equation}
    \lambda_{i} = \max_{j \in [pos_{i}+1,m]} (\mathbf{u}^{l}_{i} \cdot \mathbf{p}^{l}_{j} + \|\mathbf{u}^{r}_{i}\|\|\mathbf{p}^{r}_{j}\|),  \label{eq:ip_upper-bound}
\end{equation}
which is the maximum upper-bound of the inner product of $\mathbf{u}_i$ and $\mathbf{p}'$, where $\mathbf{p}' \in \mathbf{P}$ has not been accessed for $\mathbf{u}_i$. 
This will be used in a query phase.
Notice that the above evaluation does not compute inner products and update $A^{i}$, so it is not within the scope of $B$.
Last, for each $k \in [1, k_{max}]$, if $\mathbf{p}$ is included in the first $k$ elements of $A_{i}$, we increment $uscore_{k}(\mathbf{p})$ by one.

\subsubsection{\textbf{Analysis}}
We here analyze the space complexity, time complexity, and correctness of our pre-processing algorithm.

\begin{theorem} \label{theorem:offline}
The space and time complexities of Algorithm \ref{algo:pre-processing} are $O((n+m))d)$ and $O(nm)$, respectively. 
\end{theorem}

\noindent
\textsc{Proof.}
Algorithm \ref{algo:pre-processing} makes (i) the norm of each vector, (ii) a transformed vector for each one, (iii) $A_{i}$ for each $\mathbf{u}_{i} \in \mathbf{U}$, and (iv) $\lambda_{j}$, the value obtained in Equation (\ref{eq:ip_upper-bound}), for each $\mathbf{u}_{j} \in \mathbf{U}''$.
Therefore, we have $O(n+m)$ norm values, $O(n+m)$ $d$-dimensional transformed vectors, $O(n)$ arrays, each of which has $O(k_{max})$ space, and $O(|\mathbf{U}''|)$ values. 
Since $k_{max} = O(1)$ and $|\mathbf{U}''| \leq n$, the space complexity is $O((n+m)d)$.

Next, step (1) needs $O((n+m)d)$ and $O(m\log m)$ times to compute the norm of all vectors and sorting, respectively.
Step (2) needs $O((n+m)d^2)$ and $O((n+m)d)$ times for SVD transformation and norm computation, respectively.
By setting $B = O(nk_{max})$, step (3) needs $O(nk_{max}d + |\mathbf{U}'|\log m + |\mathbf{U}'|\log |\mathbf{U}'|) = O(n(k_{max}d + \log m + \log n))$ time.
Because Equation (\ref{eq:dynamic}) can be solved in $O(1)$ time, step (4) needs $O(|\mathbf{U}'|k_{max}d) = O(nk_{max}d)$ time.
Last, step (5) needs $O(\sum_{|\mathbf{U}''|}(m - pos_{i})) = O(nm)$ time in the worst case.
Summarizing the time complexities of steps (1)--(5) with the fact of $d \ll n,m$, the time complexity of Algorithm \ref{algo:pre-processing} is $O(nm)$.
\wsq

\begin{remark}
Theorem \ref{theorem:offline} shows that our pre-processing obtains upper-bound scores as if without inner product computations, since its time complexity does not have a factor of $d$.
This means that our pre-processing algorithm is much faster than a brute-force-like upper-bounding approach that needs $O(nmd)$ time, as $d$ is usually hundreds.
In addition, thanks to the dynamic assignment approach, many user vectors can have $\mathbf{S}_{k_{max}}(\cdot)$ in pre-processing.
This means that $|\mathbf{U}''| \ll n$, so Algorithm \ref{algo:pre-processing} practically scales better than $O(nm)$.
\end{remark}

\begin{theorem}
Algorithm \ref{algo:pre-processing} guarantees that $uscore_{k}(\mathbf{p}) \geq score_{k}(\mathbf{p})$ for each $k \in [1, k_{max}]$ and $\mathbf{p} \in \mathbf{P}$.
\end{theorem}

\noindent
\textsc{Proof.}
To prove this theorem, it is sufficient to show that whenever $\mathbf{p}$ is in $\mathbf{S}_{k}(\mathbf{u}_{i})$, $uscore_{k}(\mathbf{p})$ is incremented by using $A_{i}$.
This is trivially true for $\mathbf{u}_{i}$ such that $A_{i} = \mathbf{S}_{k_{max}}(\mathbf{u}_{i})$.
Hence, we focus on each $\mathbf{u}_{j} \in \mathbf{U}''$, i.e., $\mathbf{u}_{j}$ such that $A_{j}$ is not guaranteed to be $\mathbf{S}_{k_{max}}(\mathbf{u}_{j})$.

Assume that $\mathbf{p} \in \mathbf{S}_{k}(\mathbf{u}_{j})$ is not included in $A_{j}$.
Recall that $uscore_{k}(\mathbf{p})$ is incremented by one if $\mathbf{u}^{l}_{j} \cdot \mathbf{p}^{l} + \|\mathbf{u}^{r}_{j}\|\|\mathbf{p}^{r}\| \geq A_{j}^{k}$, see Inequality (\ref{eq:upper-bound}) and line \ref{algo:pre-processing:ub}.
From Inequality (\ref{eq:inc}) and the assumption of $\mathbf{p} \in \mathbf{S}_{k}(\mathbf{u}_{j})$, we have $\mathbf{u}^{l}_{j} \cdot \mathbf{p}^{l} + \|\mathbf{u}^{r}_{j}\|\|\mathbf{p}^{r}\| \geq \mathbf{u}_{j} \cdot \mathbf{p} \geq A_{j}^{k}$, which proves the correctness of this theorem.
\wsq

\subsection{Query Processing}
Given $N$ and $k$, we compute the top-$N$ result set $\mathbf{R}$, by exploiting the upper-bound scores obtained in pre-processing.
Before presenting our query processing algorithm, we introduce our reverse $k$-MIPS algorithm that exploits the information obtained in our pre-processing.

\subsubsection{\textbf{Incremental Reverse $k$-MIPS Algorithm.}}    \label{sec:proposal:rmips}
Recall that $\mathbf{U}''$ is a set of user vectors $\mathbf{u}_{i}$ such that $A_i$ is not guaranteed to be $\mathbf{S}_{k_{max}}(\mathbf{u}_{i})$ (i.e., its ``$k_{max}$''-MIPS result is not guaranteed).
Even in this case, $A_i$ may contain $\mathbf{S}_{k}(\mathbf{u}_{i})$ (i.e., its ``$k$''-MIPS result is guaranteed), which can be confirmed in $O(1)$ time by using $\lambda_{i}$, because 
\begin{equation}
    \lambda_{i} \geq \max_{j \in [pos_{i}+1,m]} \mathbf{u}_{i} \cdot \mathbf{p}_{j}.    \label{eq:lambda}
\end{equation}
That is, if $A^{k}_{i} \geq \lambda_{i}$, $\mathbf{u}_{i}$ has $\mathbf{S}_{k}(\mathbf{u}_{i})$.
We therefore need to focus only on each user vector that does not have $\mathbf{S}_{k}(\cdot)$.
Let $\mathbf{X}$ be a set of such user vectors.

Given an item vector $\mathbf{p}$, to obtain $score_{k}(\mathbf{p}_{j})$, we need to check whether $\mathbf{p}_{j} \in \mathbf{S}_{k}(\mathbf{u}_{i})$ for each $\mathbf{u}_{i} \in \mathbf{X}$.
If $pos_{i} \geq j$ and $\mathbf{p}_{j}$ is not included in the first $k$ elements of $A_{i}$, it is trivial that $\mathbf{p}_{j} \notin \mathbf{S}_{k}(\mathbf{u}_{i})$.
Otherwise, we check whether $\|\mathbf{u}_{i}\|\|\mathbf{p}_{j}\| > A^{k}_{i}$.
If yes, we next check whether $\mathbf{u}_{i}^{l} \cdot \mathbf{p}_{j}^{l} + \|\mathbf{u}_{i}^{r}\|\|\mathbf{p}_{j}^{r}\| > A^{k}_{i}$.
If this is still yes, we compute $\mathbf{u}_{i} \cdot \mathbf{p}_{j}$.
If and only if $\mathbf{u}_{i} \cdot \mathbf{p}_{j} > A^{k}_{i}$ and $\mathbf{u}_{i} \cdot \mathbf{p}_{j} \geq \lambda_{i}$, it is guaranteed that $\mathbf{p}_{j} \in \mathbf{S}_{k}(\mathbf{u}_{i})$.
If $\mathbf{u}_{i} \cdot \mathbf{p}_{j} > A^{k}_{i}$ but $\mathbf{u}_{i} \cdot \mathbf{p}_{j} < \lambda_{i}$, we run a linear scan of $\mathbf{P}$ from the $(pos_{i}+1)$-th item vector to obtain $\mathbf{S}_{k}(\mathbf{u}_{i})$ in the same way in pre-processing.
% In this case, we update $pos_{i}$, which is the last position of inner product computation for $\mathbf{u}_{i}$.
Finally, we return the number of user vectors in $\mathbf{X}$ such that $\mathbf{p}_{j} \in \mathbf{S}_{k}(\cdot)$.
(Once we obtain $\mathbf{S}_{k}(\mathbf{u}_{i})$ for $\mathbf{u}_{i} \in \mathbf{X}$ online, we (i) update the score of the item vectors in $\mathbf{S}_{k}(\mathbf{u}_{i})$ while avoiding duplicate counting and (ii) remove  $\mathbf{u}_{i}$ from $\mathbf{X}$.)

\begin{remark}  \label{remark:rmips}
The above approach is considered as an extended version of Simpfer \cite{amagata2021reverse} because our approach also solves the $k$-MIPS decision problem.
Simpfer is optimized for reverse $k$-MIPS for a given item vector.
That is, it does not consider that Simpfer is conducted for other item vectors $\in \mathbf{P}$ in a sequential manner.
Simply running Simpfer on different item vectors can incur duplicate linear scans for the same user vectors.
Our algorithm avoids such duplication by using $pos_{i}$, which enables incremental scan.
\end{remark}

\subsubsection{\textbf{The Algorithm}}
Now we are ready to introduce our query processing algorithm.
Algorithm \ref{algo:query} describes it, and \textsc{Inc-RMIPS} in Algorithm \ref{algo:query} represents the algorithm in Section \ref{sec:proposal:rmips}.
Our algorithm consists of two steps.

\vs
\noindent
\underline{\textbf{(1) Initialization.}}
We first initialize the score of each $\mathbf{p} \in \mathbf{P}$.
Each $\mathbf{u}_{i} \in \mathbf{U}$ such that $A^{k}_{i} \geq \lambda_{i}$ has $\mathbf{S}_{k}(\mathbf{u}_{i})$.
Hence, for each $\mathbf{p}$ in the first $k$ elements of $A_{i}$, we increment $score_{k}(\mathbf{p})$ by one.
On the other hand, if $A^{k}_{i} < \lambda_{i}$, we add $\mathbf{u}_{i}$ into a set $\mathbf{X}$.
After that, we make a copy of $\mathbf{P}$, denoted by $\mathbf{P}_{cpy}$, and sort the item vectors in $\mathbf{P}_{cpy}$ in descending order of $uscore_{k}(\cdot)$.
Note that $\mathbf{P}$ is used for reverse $k$-MIPS, which requires a norm-based sort order.

\begin{algorithm}[!t]
    \caption{\textsc{Query processing algorithm}}
    \label{algo:query}
    \DontPrintSemicolon
    \KwIn {$\mathbf{U}$, $\mathbf{P}$, $k$, and $N$}

    $\mathbf{R} \gets \varnothing$, $\tau \gets 0$, $\mathbf{X} \gets \varnothing$\;
    \ForEach {$\mathbf{u}_i \in \mathbf{U}$}{
        \eIf {$A_{i}^{k} \geq \lambda_{i}$}{ 
            \ForEach {$\mathbf{p}$ in the first $k$ elements of $A_{i}$}{
                $score_{k}(\mathbf{p}) \gets score_{k}(\mathbf{p}) + 1$
            }
        }{
            Add $\mathbf{u}_{i}$ into $\mathbf{X}$
        }
    }
    $\mathbf{P}_{cpy} \gets \mathbf{P}$\;
    Sort $\mathbf{P}_{cpy}$ in descending order of $uscore_{k}(\cdot)$\;
    \ForEach {$\mathbf{p} \in \mathbf{P}_{cpy}$}{
        \eIf {$uscore_{k}(\mathbf{p}) > \tau$}{
            $score_{k}(\mathbf{p}) \gets score_{k}(\mathbf{p}) +$ \textsc{Inc-RMIPS}$(\mathbf{p}, \mathbf{X}, \mathbf{P}, k)$\;
            Update $\mathbf{R}$ and $\tau$ if necessary
        }{
            \textbf{return} $\mathbf{R}$
        }
    }
\end{algorithm}

\vs
\noindent
\underline{\textbf{(2) Computing the top-$N$ result.}}
We compute the score of each item vector in $uscore_{k}(\cdot)$-based order.
Given an item vector $\mathbf{p}$, we update $score_{k}(\mathbf{p})$ via our reverse $k$-MIPS algorithm introduced above.
After evaluating $N$ item vectors, we obtain an intermediate result set $\mathbf{R}$ and a threshold $\tau$, the current $N$-th largest score.
Assume that we now evaluate the $(N+1)$-th item vector, say $\mathbf{p}'$.
If $uscore_{k}(\mathbf{p}') > \tau$, $\mathbf{p}'$ can be in the final result, so we need to compute $score_{k}(\mathbf{p}')$.
On the other hand, if $uscore_{k}(\mathbf{p}') \leq \tau$, we can guarantee that the current $\mathbf{R}$ is the final result.
Therefore, whenever we have $uscore_{k}(\mathbf{p}') \leq \tau$, we terminate the query processing.

\begin{remark}
Algorithm \ref{algo:query} incurs $O(nk + m\log m + nmd) = O(nmd)$ time because, in the worst case, Algorithm \ref{algo:query} needs to run $k$-MIPS for each user vector (during reverse $k$-MIPS).
However, this rarely occurs in practice, as our experimental results confirm.
Furthermore, this worst case theoretically means that our algorithm is reduced to the baseline algorithm.
This is not a negative result but a positive, because it is theoretically guaranteed that our algorithm cannot be beaten by the baseline.
% (Due to the properties in Remark \ref{remark:rmips}, our algorithm practically has less sequential scan costs than the baseline.)
\end{remark}

\section{Experiment}    \label{sec:experiment}
This section reports our experimental results.
All experiments were conducted on a Ubuntu 22.04 LTS machine with 24GB RAM and 2.8GHz Core i9-10900 CPU.

\subsection{Setting}    \label{sec:experiment:setting}
\noindent
\underline{\textbf{Dataset.}}
We used the following four commonly used real datasets.
\begin{itemize}
    \setlength{\leftskip}{-5.0mm}
    \item   Amazon-Kindle \cite{he2016ups}:
            A rating dataset for books in Amazon Kindle.
            The numbers of users and items are respectively 1,406,890 and 430,530.
    \item   Amazon-Movie \cite{he2016ups}:
            A rating dataset for movies in Amazon.
            The numbers of users and items are respectively 2,088,620 and 200,941.
    \item   MovieLens\footnote{\url{https://grouplens.org/datasets/movielens/}}:
            This is the MovieLens 25M dataset.
            The numbers of users and items are respectively 162,541 and 59,047.
    \item   Netflix\footnote{\url{https://www.cs.uic.edu/liub/Netflix-KDD-Cup-2007.html}}:
            This is a rating dataset used in Netflix Prize.
            The numbers of users and items are respectively 480,189 and 17,770.
\end{itemize}
As with the existing reverse $k$-MIPS works \cite{amagata2021reverse,amagata2023reverse,huang2023sah}, we used Matrix Factorization \cite{chin2016libmf} to generate user and item vectors.
The dimensionality of each vector was 200.

\vs
\noindent
\underline{\textbf{Competitors.}}
We compared our algorithm with the following baselines.
\begin{itemize}
    \setlength{\leftskip}{-5.0mm}
    \item   LEMP \cite{teflioudi2015lemp}:
            A state-of-the-art algorithm that runs $k$-MIPS for each user.
            We used this algorithm to obtain the score of each item vector, as introduced in Section \ref{sec:preliminary:baseline}.
    \item   FEXIPRO \cite{li2017fexipro}:
            A state-of-the-art $k$-MIPS algorithm.
            We ran this algorithm for each user vector, as with LEMP.
    \item   Simpfer \cite{amagata2021reverse}:
            A state-of-the-art reverse $k$-MIPS algorithm.
            We ran this algorithm for each item to obtain its score.
            (We did not run duplicate linear scans of $\mathbf{P}$ for the same users by using $pos_{i}$, for fair comparison.)
\end{itemize}
All algorithms were implemented in C++, compiled by g++ 11.4.0 with -O3 optimization, and single threaded. 
All evaluated algorithms return the exact answer, so our experiments focused on computational time.
We set $k_{max} = 25$ \cite{amagata2021reverse,amagata2023reverse} and $d' = 10$.

\subsection{Pre-processing Result}
We first show that our pre-processing algorithm terminates within a reasonable time.
Table \ref{tab:pre-processing} shows the result.
Even when we have million-scale users (e.g., Amazon-Kindle and Amazon-Movie), our pre-processing needs only about 10 minutes.
Recall that this pre-processing is done once (as long as $k_{max}$ does not increase), so the pre-processing time is reasonable.
Thanks to this offline step, our query processing is significantly efficient, e.g., its time is usually less than 1 second, enabling an interactive response, as seen later.

\begin{table}[t]
    \centering
    \caption{Pre-processing time [sec]}
    \label{tab:pre-processing}
    \begin{tabular}{cccc}   \toprule
        Amazon-Kindle   & Amazon-Movie  & MovieLens & Netflix   \\ \midrule
        743             & 657           & 24        & 41        \\ \bottomrule
    \end{tabular}
\end{table}

\subsection{Query Processing Result}
\begin{figure}[!t]
    \begin{center}
        \subfigure[Amazon-Kindle]{%
		\includegraphics[width=0.48\linewidth]{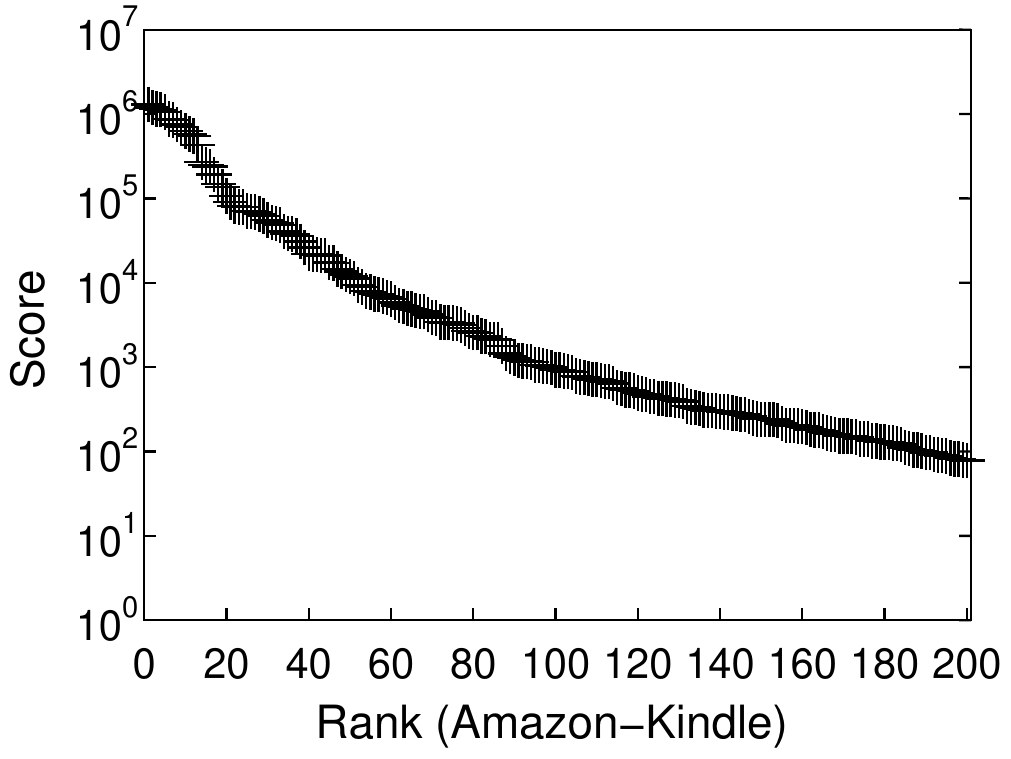}  \label{fig:amazon-k-score}}
        \subfigure[Amazon-Movie]{%
		\includegraphics[width=0.48\linewidth]{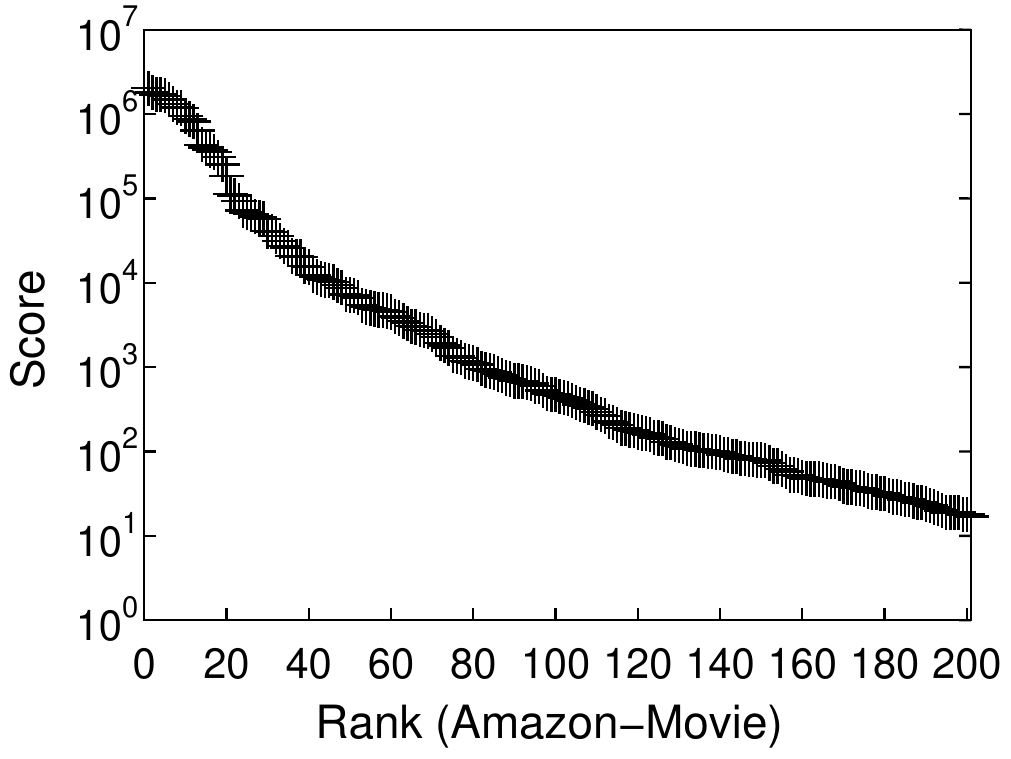}  \label{fig:amazon-m-score}}
	\subfigure[MovieLens]{%
		\includegraphics[width=0.48\linewidth]{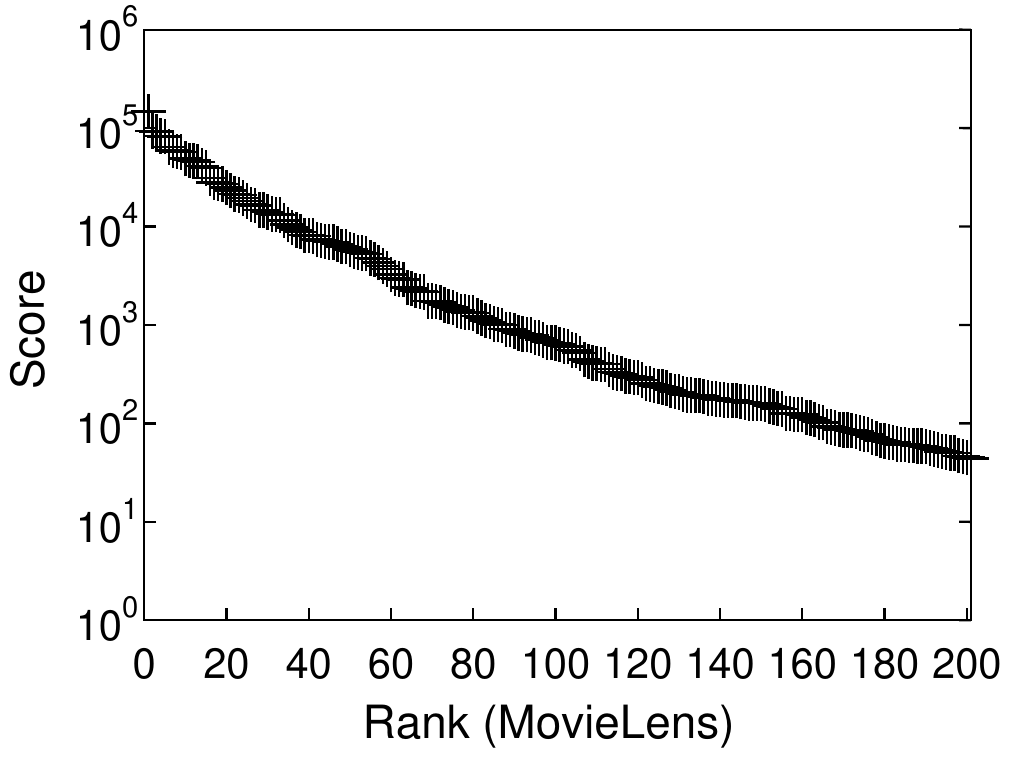} \label{fig:movielens-score}}
        \subfigure[Netflix]{%
		\includegraphics[width=0.48\linewidth]{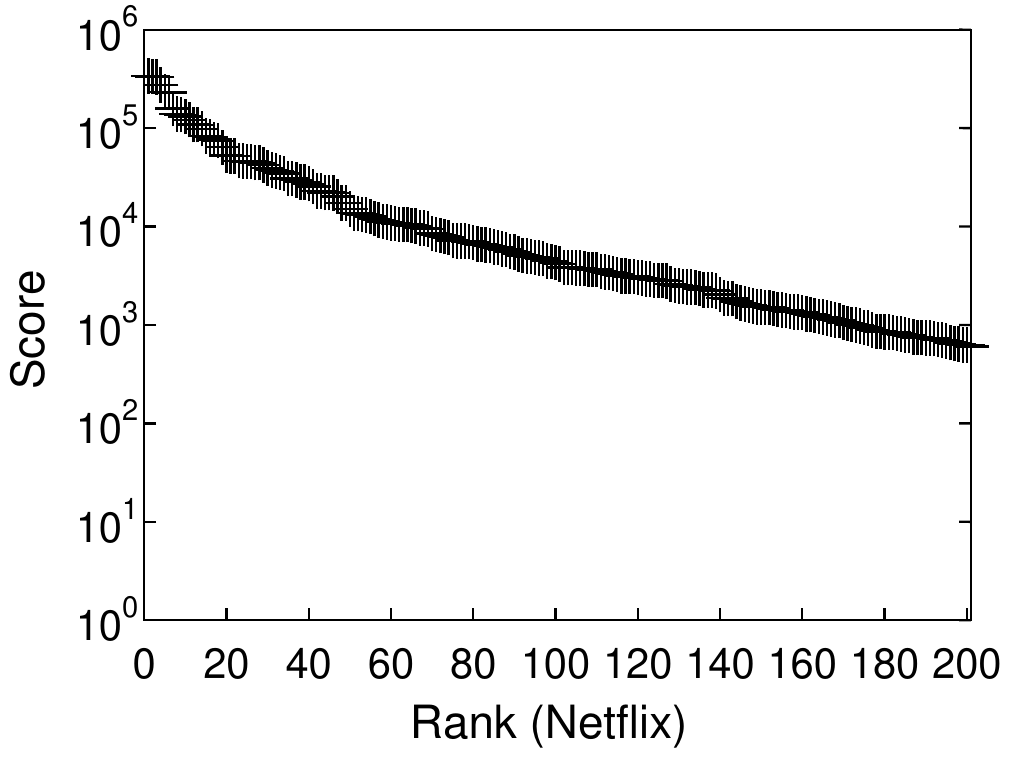}   \label{fig:netflix-score}}
        \caption{Score distribution of each dataset}
        \label{fig:score}
    \end{center}
\end{figure}

\noindent
\textbf{Top-$N$ result can attract many users.}
First, to confirm that our problem outputs meaningful results, we visualize the score distribution of each dataset ($k = 10$) by using rank order (up to 200).
Figure \ref{fig:score} illustrates the score distributions.
We see that top-ranked item vectors have large scores, meaning that they are contained in the $k$-MIPS results of many users.
Our problem successfully outputs these items.
Subsequent applications can use such potentially popular items to predict which items can be popular next and to develop new items. 

Hereinafter, we focus on query processing time.
By default, we set $k = 10$ and $N = 20$.

\vs
\noindent
\textbf{Effectiveness of our dynamic budget assignment.}
Next, we compare our algorithm with three different budget assignment methods to demonstrate that our algorithm effectively uses the budget and obtains tight upper-bound scores.
We used three polynomial methods: uniform, linear, and quadratic.
Uniform assigns the budget equally, i.e., each user vector can access $O(k_{max})$ item vectors.
Linear and quadratic are variants of Equation (\ref{eq:f_x}), and they respectively use linear and quadratic functions to assign the budget.
They share Algorithm \ref{algo:query} for query processing.

Table \ref{tab:function} shows the comparison result.
Uniform shows the worst query processing time, whereas our algorithm is the fastest.
This result clarifies that, with the same budget, our budget assignment approach provides tighter upper-bound scores than the others.
In particular, the result on Amazon-Kindle highlights our effectiveness: our algorithm needs only 0.2 seconds, whereas the others need more than 30 seconds.
Although Figure \ref{fig:function} has already suggested that our exponential function is appropriate, this empirical result further supports this intuition.

\begin{table}[t]
    \centering
    \caption{Query processing time [sec]}
    \label{tab:function}
    \begin{tabular}{ccccc}   \toprule
        Dataset         & Ours          & Uniform   & Linear & Quadratic    \\ \midrule
        Amazon-Kindle   & \textbf{0.23} & 166.75    & 58.23  & 35.24        \\
        Amazon-Movie    & \textbf{0.22} & 158.56    & 0.96   & 0.23         \\
        MovieLens       & \textbf{0.07} & 0.65      & 0.28   & 0.21         \\
        Netflix         & \textbf{0.10} & 0.94      & 0.48   & 0.31         \\
        \bottomrule
    \end{tabular}
\end{table}

\begin{figure}[!t]
    \begin{center}
        \subfigure[Amazon-Kindle]{%
		\includegraphics[width=0.48\linewidth]{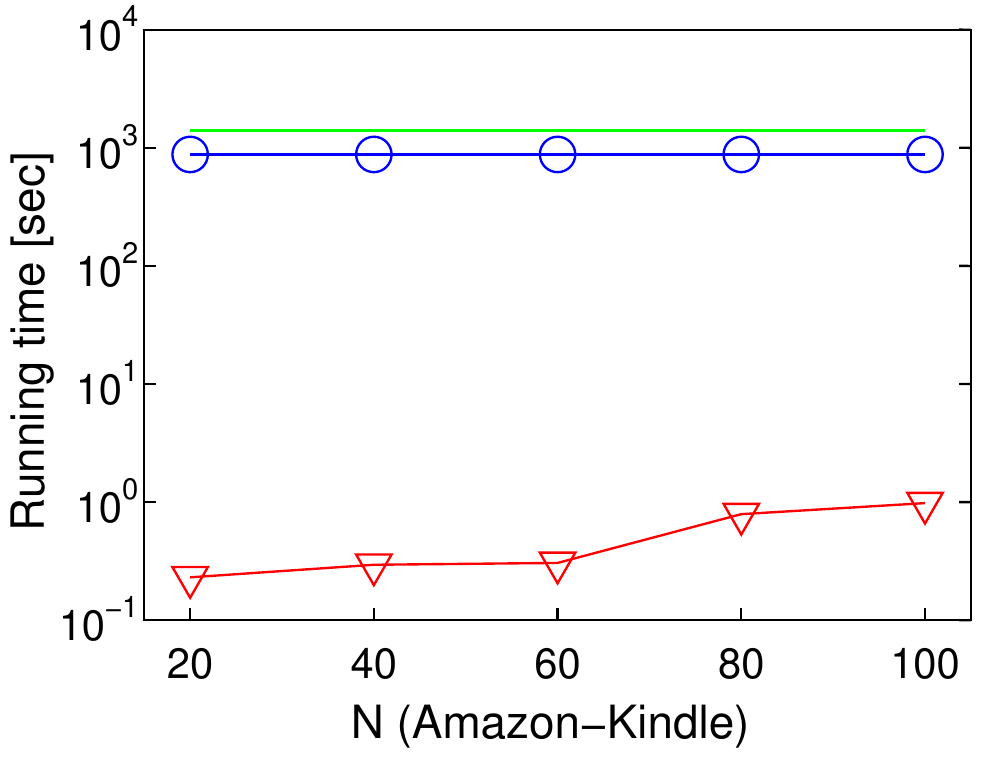}  \label{fig:amazon-k-N}}
        \subfigure[Amazon-Movie]{%
		\includegraphics[width=0.48\linewidth]{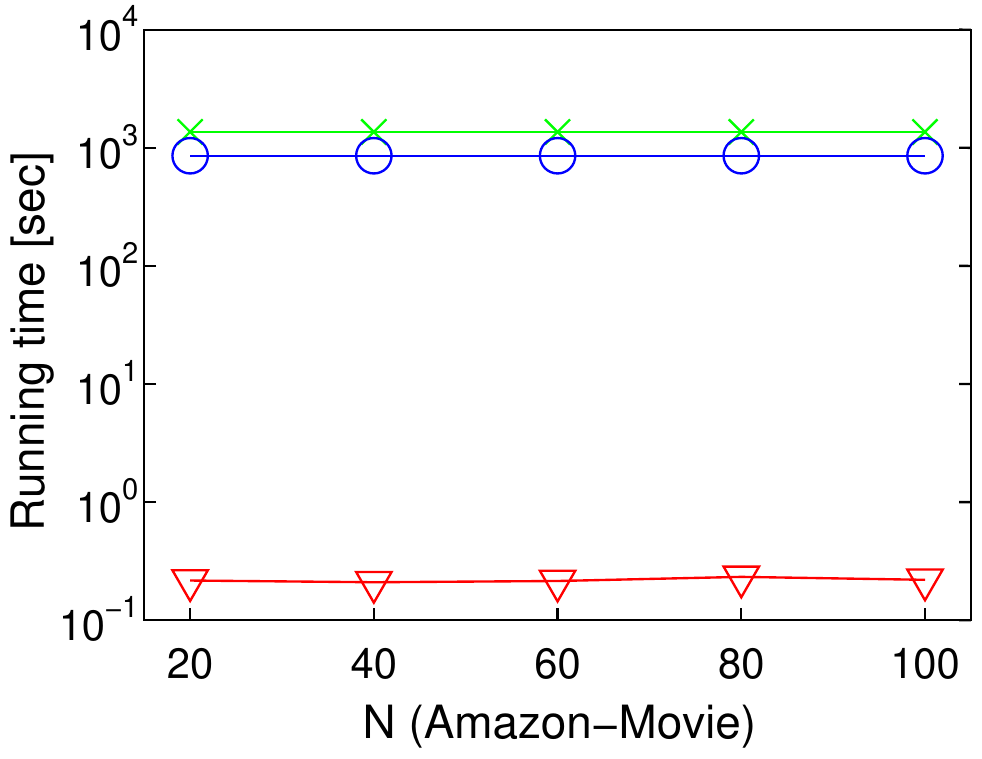}  \label{fig:amazon-m-N}}
	\subfigure[MovieLens]{%
		\includegraphics[width=0.48\linewidth]{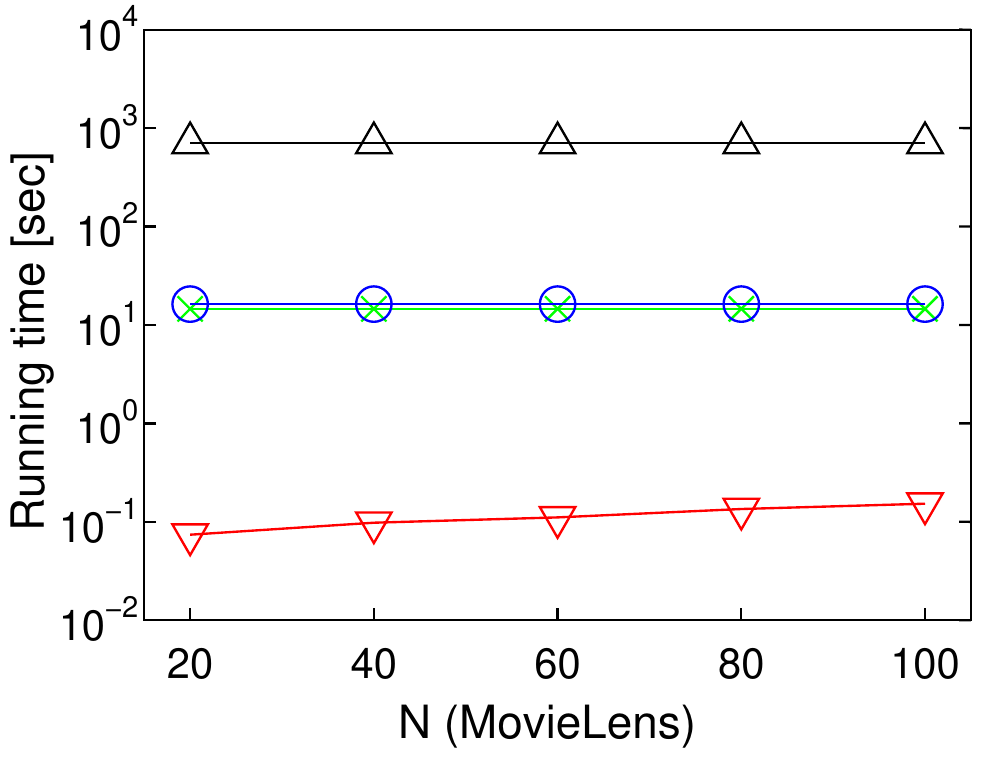} \label{fig:movielens-N}}
        \subfigure[Netflix]{%
		\includegraphics[width=0.48\linewidth]{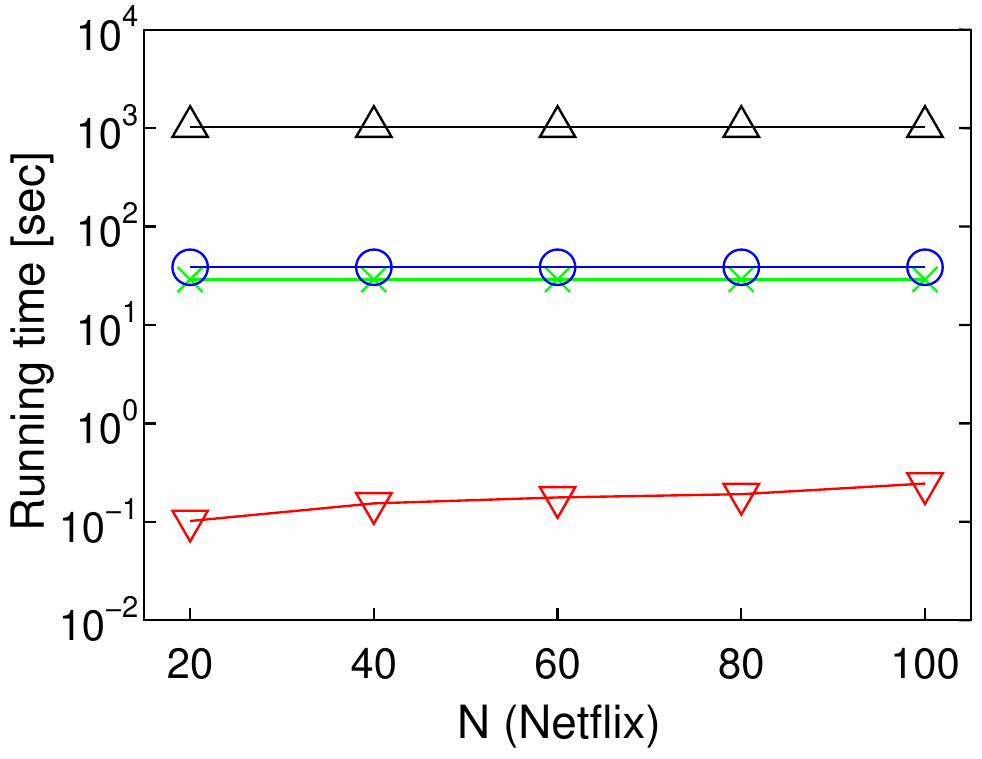}   \label{fig:netflix-N}}
        \caption{Impact of $N$:
        ``\textcolor{green}{$\times$}'' shows \textcolor{green}{LEMP}, ``\textcolor{blue}{$\circ$}'' shows \textcolor{blue}{FEXIPRO}, ``$\triangle$'' shows Simpfer, and ``\textcolor{red}{$\triangledown$}'' shows \textcolor{red}{Ours}.}
        \label{fig:N}
    \end{center}
\end{figure}

\vs
\noindent
\textbf{Impact of $N$.}
To demonstrate that our upper-bounding approach functions significantly well, we compared our algorithm with the competitors introduced in Section \ref{sec:experiment:setting} while varying some parameters.
To start with, we investigate the impact of the result size $N$, and Figure \ref{fig:N} shows the result.
(The running time of Simpfer exceeded 1 hour on Amazon-Kindle and Amazon-Movie, so we stopped the experiments.)

We have three main observations here.
First, our algorithm is significantly faster than the competitors and is the only algorithm enabling an interactive response time.
For example, when $N = 20$, our algorithm is about 3800, 3900, 200, and 280 times faster than the best competitors on Amazon-Kindle, Amazon-Movie, MovieLens, and Netflix, respectively.
This result clearly demonstrates that our upper-bounding approach functions well.
Second, the competitors are not affected by $N$.
This result is reasonable, since they compute (reverse) $k$-MIPS results for all users (items) regardless of $N$.
Last, the running time of our algorithm (slightly) increases as $N$ increases.
Since our algorithm needs to run at least $N$ incremental reverse $k$-MIPS, this result is also reasonable.
We found that the running time of our algorithm is empirically (sub-)linear to $N$.
Therefore, our algorithm is much faster than the competitors even when $N$ is large (e.g., 100).

\begin{figure}[!t]
    \begin{center}
        \subfigure[Amazon-Kindle]{%
		\includegraphics[width=0.48\linewidth]{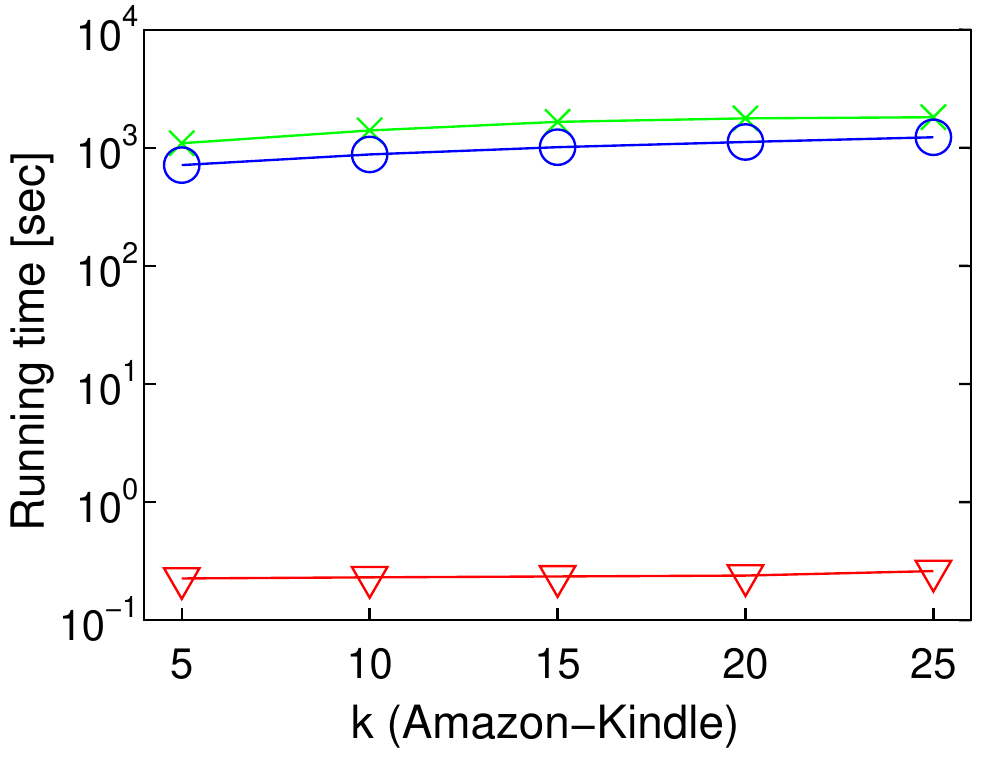}  \label{fig:amazon-k-k}}
        \subfigure[Amazon-Movie]{%
		\includegraphics[width=0.48\linewidth]{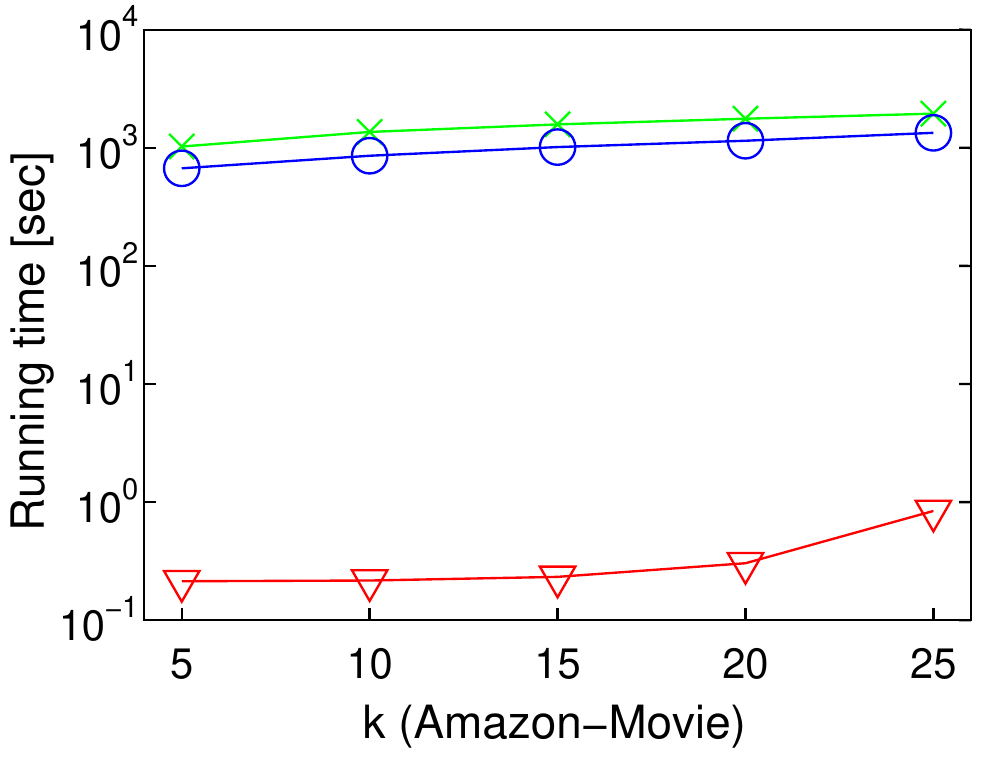}  \label{fig:amazon-m-k}}
	\subfigure[MovieLens]{%
		\includegraphics[width=0.48\linewidth]{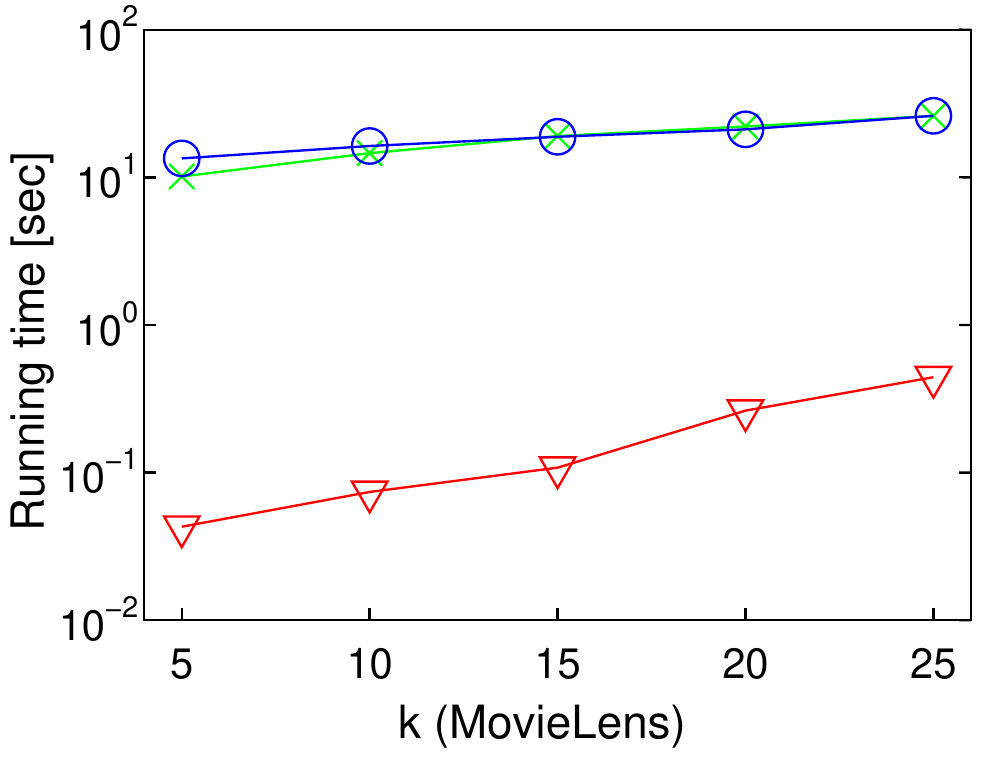} \label{fig:movielens-k}}
        \subfigure[Netflix]{%
		\includegraphics[width=0.48\linewidth]{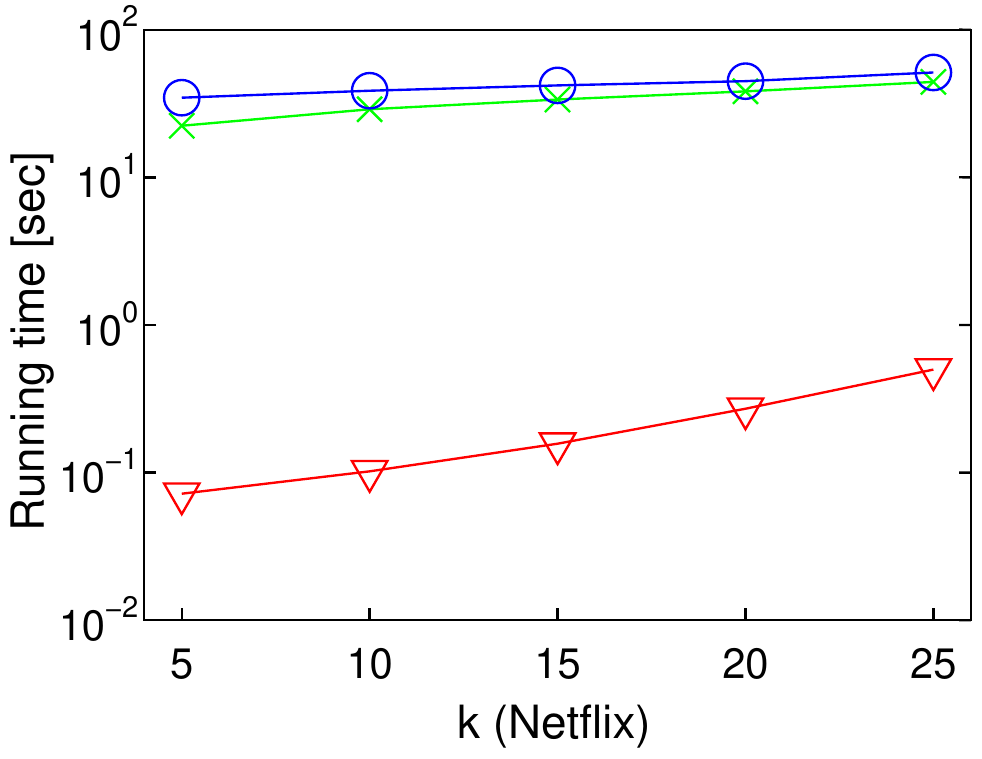}   \label{fig:netflix-k}}
        \caption{Impact of $k$:
        ``\textcolor{green}{$\times$}'' shows \textcolor{green}{LEMP}, ``\textcolor{blue}{$\circ$}'' shows \textcolor{blue}{FEXIPRO}, and ``\textcolor{red}{$\triangledown$}'' shows \textcolor{red}{Ours}.}
        \label{fig:k}
    \end{center}
\end{figure}

\vs
\noindent
\textbf{Impact of $k$.}
Figure \ref{fig:k} shows the results of experiments with varying $k$.
We do not show the result of Simpfer from now, since it is much slower than the others, as shown in Figure \ref{fig:N}.
All algorithms need longer running time as $k$ increases.
This is a trivial result for $k$-MIPS algorithms (i.e., LEMP and FEXIPRO) since they need to access more item vectors to compute the $k$-MIPS results for larger $k$.
Similarly, for the reverse $k$-MIPS algorithm (i.e., Simpfer), the computational cost of solving the decision version of the $k$-MIPS problem increases as $k$ increases.
Our algorithm also has the same case as that of Simpfer.
Because as $k$ increases, more user vectors do not have $\mathbf{S}_{k}(\cdot)$, our algorithm needs to compute the exact scores of more item vectors, which involves the $k$-MIPS decision problem.

\begin{figure}[!t]
    \begin{center}
        \subfigure[Amazon-Kindle]{%
		\includegraphics[width=0.48\linewidth]{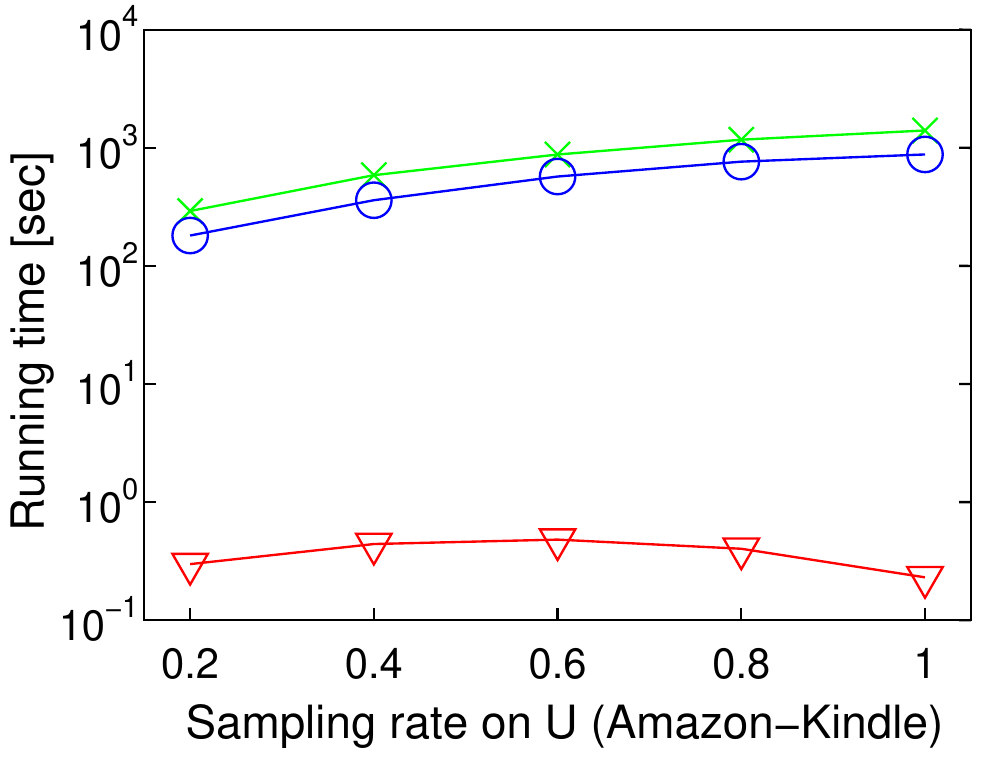}  \label{fig:amazon-k-n}}
        \subfigure[Amazon-Movie]{%
		\includegraphics[width=0.48\linewidth]{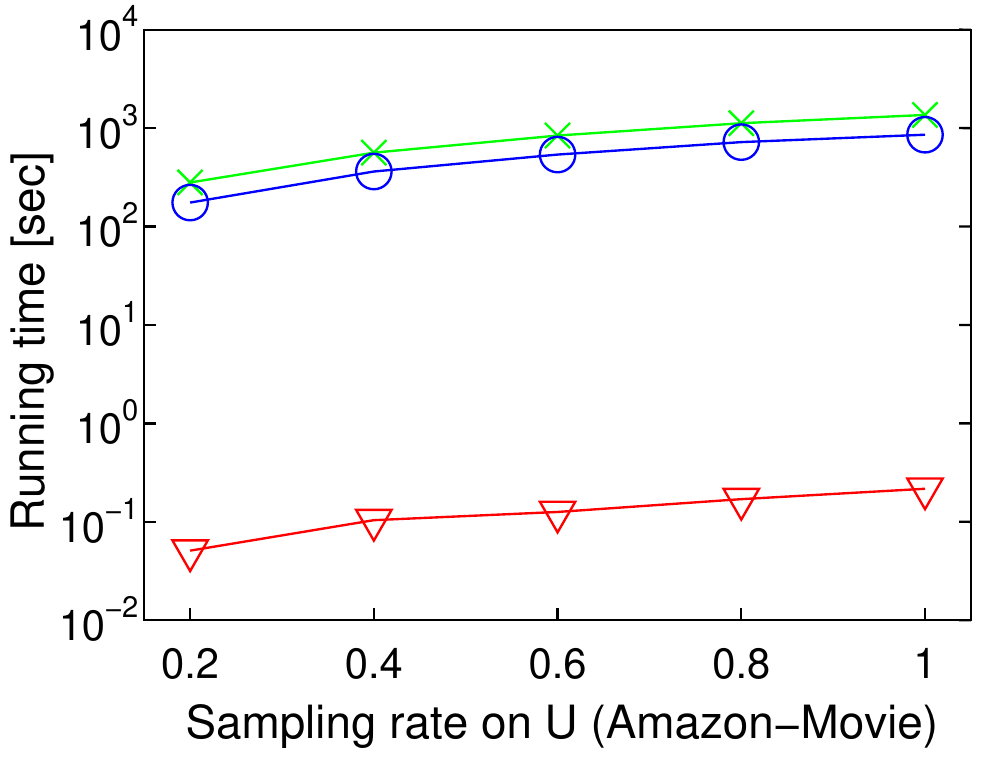}  \label{fig:amazon-m-n}}
	\subfigure[MovieLens]{%
		\includegraphics[width=0.48\linewidth]{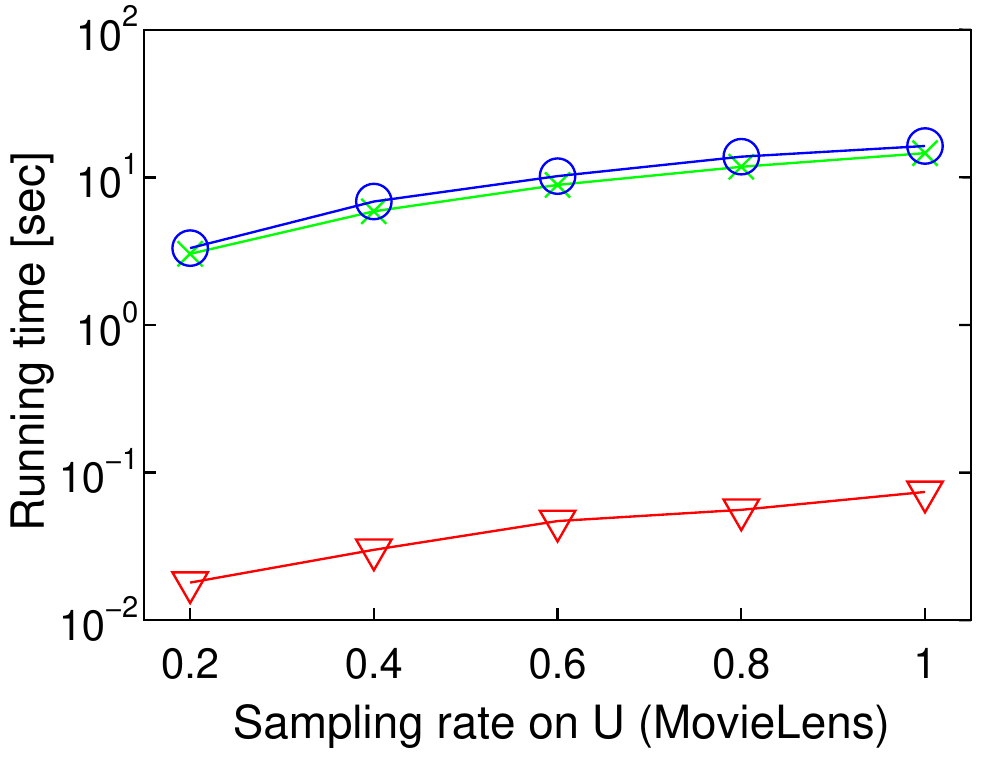} \label{fig:movielens-n}}
        \subfigure[Netflix]{%
		\includegraphics[width=0.48\linewidth]{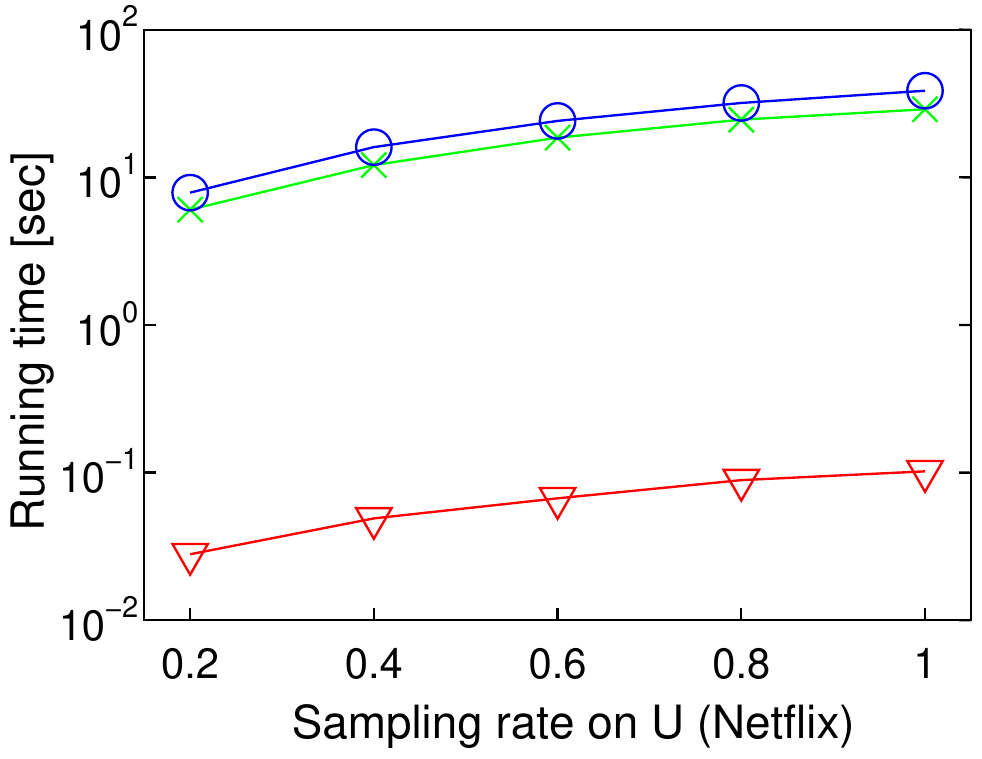}   \label{fig:netflix-n}}
        \caption{Impact of $|\mathbf{U}|$:
        ``\textcolor{green}{$\times$}'' shows \textcolor{green}{LEMP}, ``\textcolor{blue}{$\circ$}'' shows \textcolor{blue}{FEXIPRO}, and ``\textcolor{red}{$\triangledown$}'' shows \textcolor{red}{Ours}.}
        \label{fig:n}
    \end{center}
\end{figure}

\vs
\noindent
\textbf{Impact of $|\mathbf{U}|$.}
Next, to investigate the scalability to the number of users, we conducted experiments with varying the number of user vectors in $\mathbf{U}$ by random sampling.
Figure \ref{fig:n} exhibits the result.
Normally, as we have more user vectors, the running time of each algorithm increases.
This is a natural result for the competitors because their time complexities are linear to $n$.
This is also true for our algorithm, but we have an exception, i.e., the result on Amazon-Kindle (Figure \ref{fig:amazon-k-n}).
Since the efficiency of our algorithm is affected by the tightness of upper-bound scores and the score distribution of item vectors, sometimes their impacts are more weighted than that of $n$.
The result in Figure \ref{fig:amazon-k-n} showcases this observation.

\begin{figure}[!t]
    \begin{center}
        \subfigure[Amazon-Kindle]{%
		\includegraphics[width=0.48\linewidth]{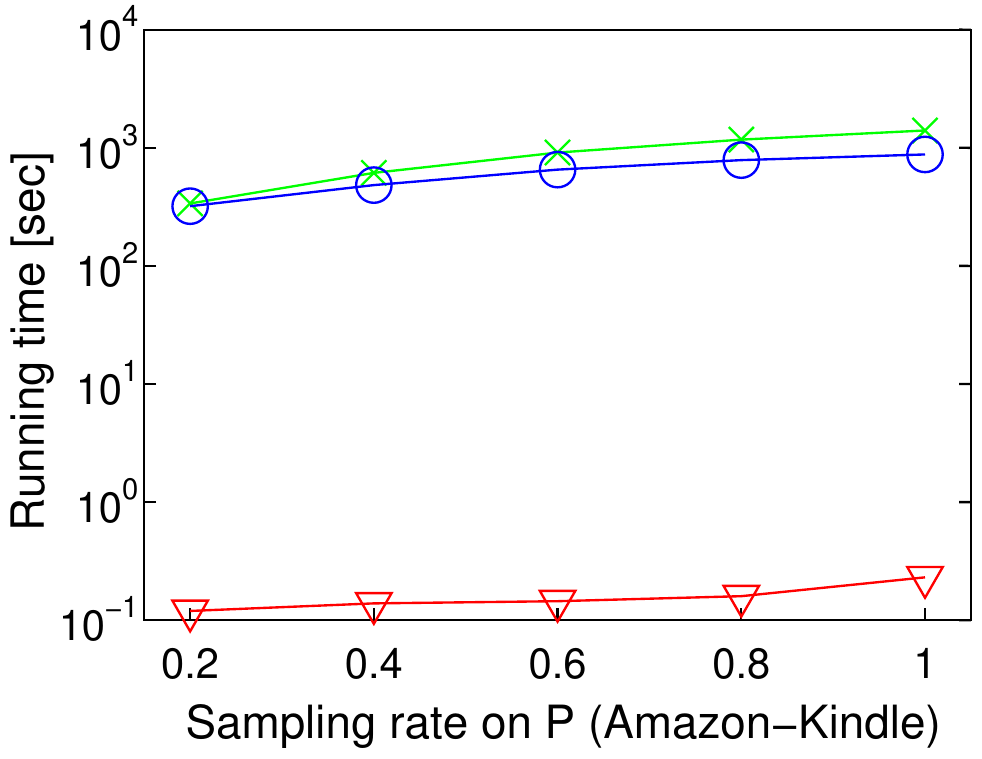}  \label{fig:amazon-k-m}}
        \subfigure[Amazon-Movie]{%
		\includegraphics[width=0.48\linewidth]{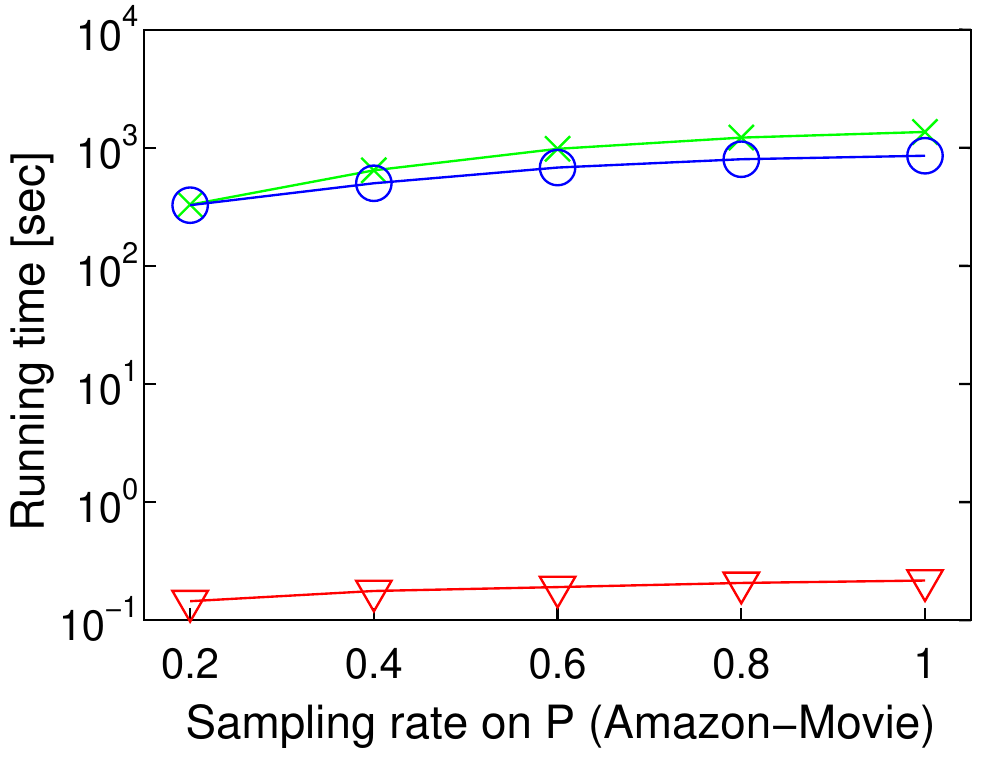}  \label{fig:amazon-m-m}}
	\subfigure[MovieLens]{%
		\includegraphics[width=0.48\linewidth]{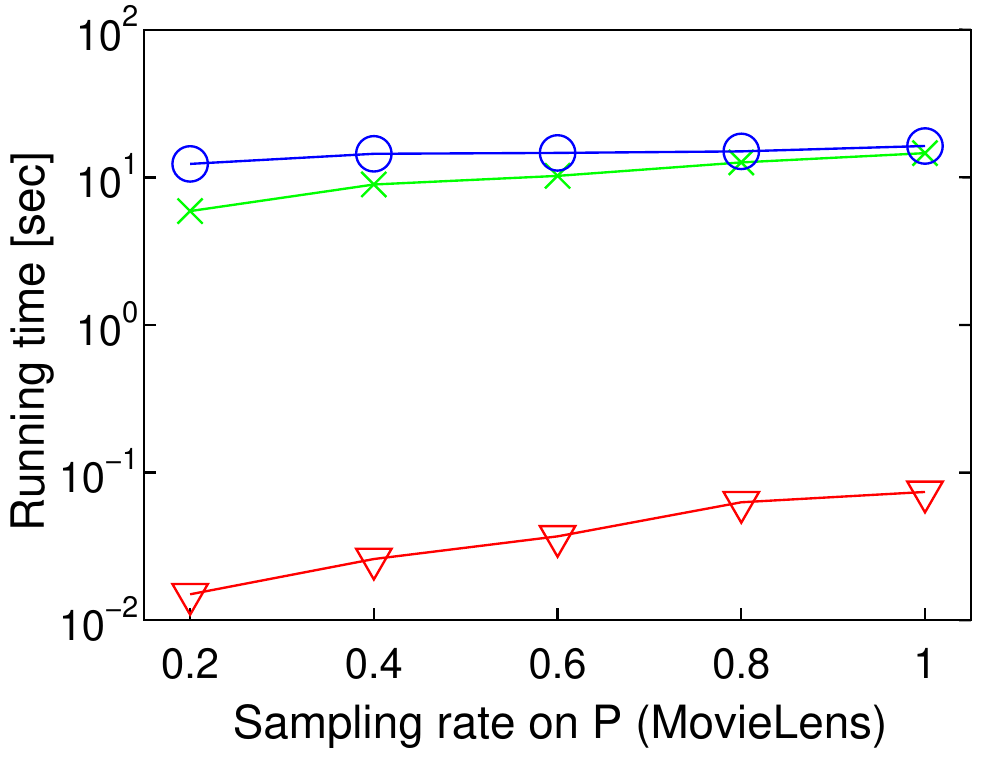} \label{fig:movielens-m}}
        \subfigure[Netflix]{%
		\includegraphics[width=0.48\linewidth]{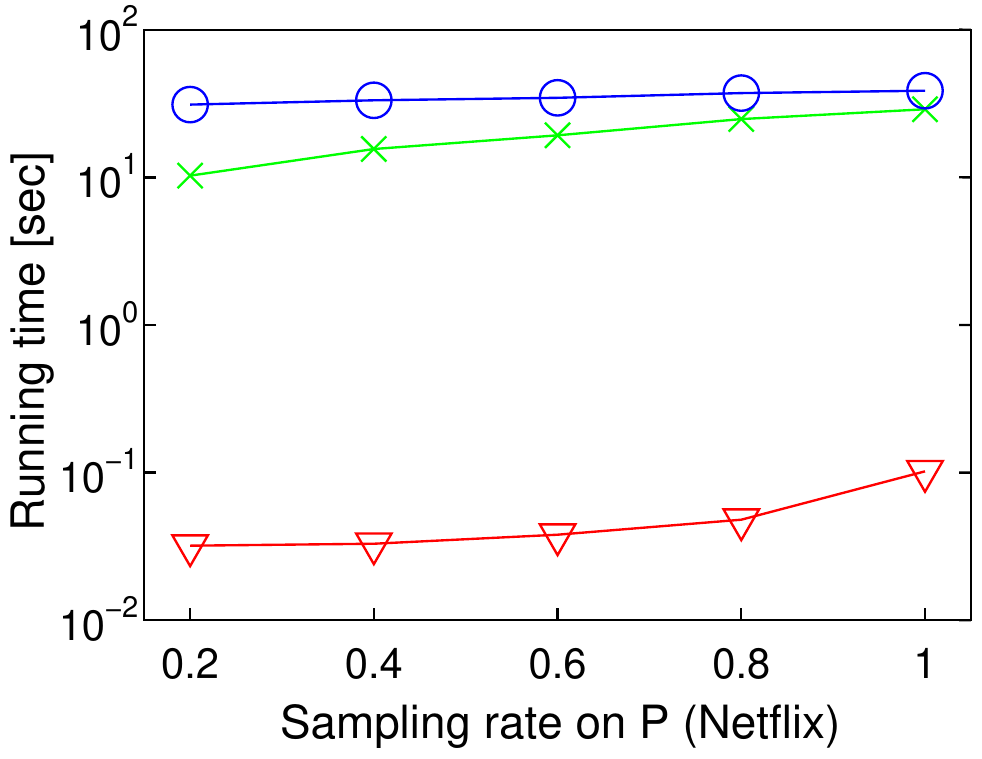}   \label{fig:netflix-m}}
        \caption{Impact of $|\mathbf{P}|$:
        ``\textcolor{green}{$\times$}'' shows \textcolor{green}{LEMP}, ``\textcolor{blue}{$\circ$}'' shows \textcolor{blue}{FEXIPRO}, and ``\textcolor{red}{$\triangledown$}'' shows \textcolor{red}{Ours}.}
        \label{fig:m}
    \end{center}
\end{figure}

\vs
\noindent
\textbf{Impact of $|\mathbf{P}|$.}
Last, we conducted experiments with varying the number of item vectors in $\mathbf{P}$ by random sampling to see the scalability to $m$.
Figure \ref{fig:m} shows the result, which is similar to the result in Figure \ref{fig:n}.
As for our algorithm, comparing the cases with sampling rates of 0.2 and 1.0, the time differences are usually less than 100 milliseconds.
This result highlights the scalability of our algorithm to the number of items.

\section{Conclusion}    \label{sec:conclusion}
This paper formulated a new problem that exploits the reverse $k$-MIPS result.
In this problem, each item vector $\mathbf{p}$ has a score, which is the cardinality of the reverse $k$-MIPS result for $\mathbf{p}$.
Then, it searches for $N$ item vectors with the largest score.
A baseline algorithm for this problem, which employs an existing $k$-MIPS or reverse $k$-MIPS algorithm, is computationally expensive.
We therefore propose an upper-bounding-based fast and exact algorithm for this problem.
We showed that our offline and online algorithms are both efficient. 
Our experimental results demonstrate that our algorithm is much faster than competitors.

\begin{acks}
This work was partially supported by AIP Acceleration Research JPMJCR23U2.
\end{acks}

\bibliographystyle{ACM-Reference-Format}
\bibliography{sigproc}

\end{document}